\documentclass[peerreview,letterpaper]{IEEEtran}
\usepackage{hyperref} %
\usepackage%
{hyperref} %
\hypersetup{
    colorlinks,
    linkcolor={blue!80!black},%
    citecolor={green!30!black},
    urlcolor={blue!80!black}
}

\usepackage{balance}

\usepackage{amsmath,amssymb,amsfonts}
\usepackage{amsmath}
\usepackage{bbm}
\usepackage{nicefrac}
\usepackage{amsthm}

\usepackage{algorithmic}

\usepackage{textcomp}
\def\BibTeX{{\rm B\kern-.05em{\sc i\kern-.025em b}\kern-.08em
    T\kern-.1667em\lower.7ex\hbox{E}\kern-.125emX}}

\usepackage{physics}
\usepackage{siunitx}
\AtBeginDocument{\RenewCommandCopy\qty\SI}
\ExplSyntaxOn
\msg_redirect_name:nnn { siunitx } { physics-pkg } { none }
\ExplSyntaxOff

\usepackage{graphicx}
\usepackage{xcolor}

\usepackage{comment}

\renewcommand{\trace}{\mathrm{Tr}}
\AtBeginDocument{\RenewCommandCopy\qty\SI}

\theoremstyle{remark}	\newtheorem{theorem}{Theorem}
\theoremstyle{remark}	
\theoremstyle{remark}	\newtheorem{corollary}[theorem]{Corollary}
\theoremstyle{remark}	
\theoremstyle{remark} \newtheorem{definition}{Definition}
\theoremstyle{remark} \newtheorem{remark}{Remark}
\theoremstyle{remark} \newtheorem{example}{Example}

\newcommand{\identity}{\mathbbm{1}}
\newcommand{\markovC}[1]{%
    \begin{tikzpicture}[#1]%
    \draw (0,0.3ex) -- (1ex,0.3ex);%
    \draw (0.5ex,0.3ex) circle (0.2ex);
    \draw[white] (0.2ex,0) -- (0.5ex,0);%
    \end{tikzpicture}%
}
\newcommand{\Cbar}{\markovC{scale=2}}

\usepackage[square,numbers,sort&compress]{natbib}
 \usepackage[utf8]{inputenc}
\usepackage{mathrsfs}

\usepackage{accents}
\newlength{\dhatheight}

\usepackage{lscape}

\usepackage{tikz}
\usetikzlibrary{positioning}
\usetikzlibrary{decorations.markings}
\usetikzlibrary{arrows}
\usetikzlibrary{arrows.meta}
\usepackage[free-standing-units]{siunitx}
\usepackage[siunitx, RPvoltages]{circuitikz}

\makeatletter
\ctikzset{bipoles/my switch/height/.initial=.5}
\ctikzset{bipoles/my switch/width/.initial=.50}
\pgfcircdeclarebipole{}{}{myswitch}{%
\ctikzvalof{bipoles/my switch/height}}{\ctikzvalof{bipoles/my switch/width}}{
        \pgfsetlinewidth{\pgfkeysvalueof{/tikz/circuitikz/bipoles/thickness}\pgfstartlinewidth}
        \pgfpathmoveto{\pgfpoint{\pgf@circ@res@left}{0pt}}
        \pgfpathlineto{\pgfpoint{\pgf@circ@res@right}{.75\pgf@circ@res@up}}
        \pgfusepath{draw}
        \pgfsetlinewidth{\pgfstartlinewidth}        
        \pgftransformshift{\pgfpoint{\pgf@circ@res@left}{0pt}}
        \pgfnode{ocirc}{center}{}{}{\pgfusepath{draw}}
        \pgftransformshift{\pgfpoint{2\pgf@circ@res@right}{0pt}}
        \pgfnode{ocirc}{center}{}{}{\pgfusepath{draw}}
}
\def\pgf@circ@myswitch@path#1{\pgf@circ@bipole@path{myswitch}{#1}}
\compattikzset{my switch/.style = {\circuitikzbasekey, 
/tikz/to path=\pgf@circ@myswitch@path}}

\usepackage{xcolor}
\definecolor{mypurple}{rgb}{1,0,1}
	\definecolor{apricot}{rgb}{0.98, 0.81, 0.69}
		\definecolor{azure}{rgb}{0.0, 0.5, 1.0}
			\definecolor{darkmidnightblue}{rgb}{0.0, 0.2, 0.4}
				\definecolor{tearose}{rgb}{0.97, 0.51, 0.47}
					\definecolor{teagreen}{rgb}{0.82, 0.94, 0.75}
						\definecolor{indigo}{rgb}{0.29, 0.0, 0.51}
							\definecolor{tyrianpurple}{rgb}{0.4, 0.01, 0.24}

\tikzstyle{myedgestyle} = [-triangle 60]
\tikzstyle{block} = [draw, shape=rectangle, minimum height=3cm, minimum width=3cm, node distance=4cm, line width=0.5pt]
\tikzstyle{sum} = [draw, shape=circle, node distance=4cm, line width=0.5pt, minimum width=2.em]
\tikzstyle{mult} = [draw, shape=circle, node distance=3cm, line width=0.5pt, minimum width=2.em]
\tikzstyle{branch}=[fill,shape=circle,minimum size=4cm,inner sep=0pt]

\allowdisplaybreaks %

\title{Entanglement Coordination Rates in Multi-User Networks}
\author{%
    \IEEEauthorblockN{Hosen Nator and Uzi Pereg} \\
    \IEEEauthorblockA{\normalsize Electrical and Computer Engineering and Helen Diller Quantum Center, Technion 
   }}

\begin{document}

\maketitle
\begin{abstract}
The optimal coordination rates are determined in three primary settings
of multi-user quantum networks, thus characterizing the minimal resources required in order to simulate a joint quantum state among multiple parties. 
We study the following models:
(1) a cascade network with rate-limited entanglement, (2) 
a broadcast network, which consists of a single sender and two receivers, (3)  
a multiple-access network with two senders and a single receiver. We establish the necessary and sufficient conditions on the asymptotically-achievable communication and entanglement rates in each setting. The examples demonstrate that coordination of entanglement and coordination of separable correlations behave differently. At last, we show the implications of our results on nonlocal games with quantum strategies.
\end{abstract}


\section{Introduction}
\label{Sec: introduction}
\begin{figure}[b]
\caption{Coordination in a network that consists of  
$N=7$ nodes. Some of the links are classical, while 
others are quantum. 
For instance, Node~2 sends classical bits to Node~3 at a limited rate $R_{2,3}$. Whereas, Node~6 sends qubits to Node~5 at a quantum rate $Q_{6,5}$. 
}
\center
\includegraphics[scale=0.85,trim={5.3cm 0 5.5cm 0}]
{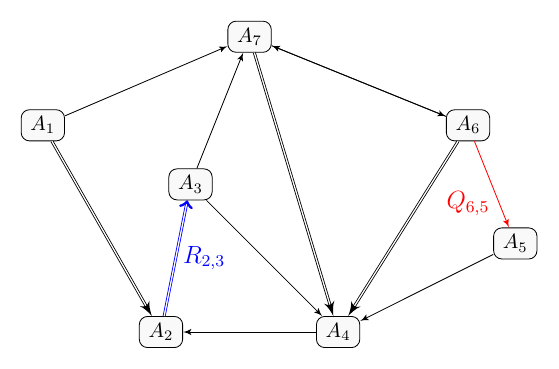} 
\label{Figure: Coordinaiton network}
\end{figure}

State distribution and coordination are important in quantum communication \cite{HsiehWilde:10p1}, %
computation \cite{matera2016coherent%
}, and cryptography \cite{liu2013experimental}. %
The quantum coordination problem can be described as follows. Consider a network that consists of $N$ nodes, where Node $i$ can perform an encoding operation $\mathcal{E}_i$ on a quantum system $A_i$, and its state should be in a certain correlation with the rest of the network nodes. An example is shown in  Figure~\ref{Figure: Coordinaiton network}.
The objective is to \emph{simulate} a specific joint state $\omega_{A_1, A_2, \dots , A_N}$, i.e., a noisy correlation.  The objective is to simulate a specific joint state $\omega_{A_1, A_2, \dots , A_N}$. Some of the nodes are not free to choose their encoding operation, but rather have their quantum state dictated by Nature via a physical process. 
Node $i$ can send qubits to node $j$ via a quantum channel at a limited communication rate ${Q_{i,j}}$. 
In addition, the nodes may share limited entanglement resources, prior to their communication.
The optimal performance of the communication network is characterized by the quantum communication rates $Q_{i,j}$ that are necessary and sufficient for simulating the desired quantum correlation. 

Instances of the network coordination problem include channel/source simulation \cite{%
channelsimulation2024,
berta2013entanglement,
wilde2018entanglement,
cheng2023quantum,george2023one,allah2024quantum,chitambar2022communication},
state merging \cite{bjelakovic2013universal,horodecki2005partial,horodecki2007quantum},
state redistribution \cite{devetak2008exact,Yard_Devetak_2009,luo2009channel,berta2016smooth},
entanglement dilution \cite{hayden2003communication,harrow2004tight,kumagai2013entanglement
},
randomness extraction \cite{6670761,
tahmasbi2020steganography},
source coding \cite{goldfeld2014ahlswede,kramer2001quantum,Compressing_mixed_state_sources_2002, 
Rate_limited_source_coding_2023},  
and many others.
Recently, we have also  considered 
   networks with classical communication links \cite{NatorPereg_CQ_No_ITW_arXiv}.

\begin{figure}[t]
    \centering
    \begin{minipage}[b]{0.45\textwidth}
        \centering
        \includegraphics[scale=0.74,trim={6cm 0 5cm 0}]{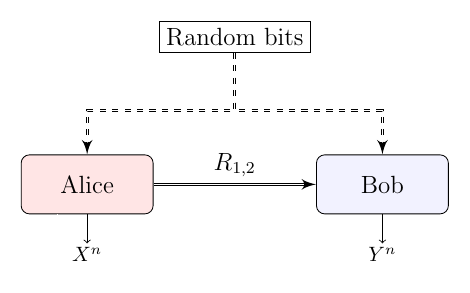}
    
\caption{Classical two-node network 
}
        \label{Group_Classical_two_nodes_network}
    \end{minipage}
    \hfill
    \begin{minipage}[b]{0.45\textwidth}
        \centering
        \includegraphics[scale=0.74,trim={7cm 0 7cm -0.5cm}]{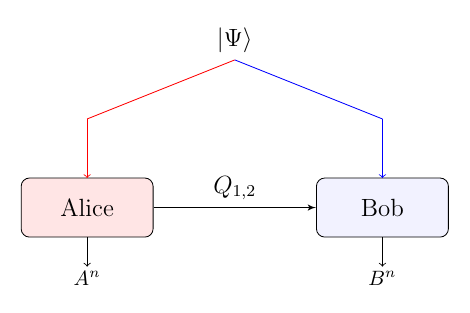}
       \caption{Quantum two-node network %
       }
        \label{Group_Quantum_two_nodes_network}
    \end{minipage}  
\end{figure} 

\paragraph*{Two-node classical coordination}
In classical coordination, the goal is to simulate a prescribed joint probability distribution.
In the basic two-node network, described in Figure~\ref{Group_Classical_two_nodes_network}, the simulation of a given joint distribution $p_{XY}$ is possible if and only if the \emph{classical} communication rate $R_{1,2}$ is above  
Wyner's common information \cite{wyner1975common}, defined as:
\begin{align}
C(X;Y) \triangleq \min
I(U;XY)\,,
\end{align}
where the minimum is taken over all auxiliary variables $U$ that satisfy the Markov relation $X \Cbar U \Cbar Y$, and
$I(U;XY)$ is the mutual information between $U$ and the pair $(X,Y)$.
In the classical model,
one may also consider the case where
the nodes share classical correlation resources, a priori, in the form of
common randomness, i.e., a sequence of
pre-shared  random bits.
Given a sufficient amount of pre-shared common randomness between the nodes,  the desired distribution
can be simulated if and only if the classical communication rate is above
the mutual information, i.e.,  $R_{1,2}\geq I(X;Y)$ \cite{bennett2002entanglement}.

\paragraph*{Two-node quantum coordination}
In the quantum setting,   the goal is to simulate a prescribed joint state. A bipartite state $\omega_{AB}$ can be simulated  
 if and only if the quantum communication rate is above the von Neumann entropy \cite{harrow2004tight}, i.e.,  
  \begin{align}
 Q_{1,2}\geq H(\omega_B)\,,
 \end{align}
 with $H(\rho)=-\trace(\rho\log_2(\rho))$,
 where $\omega_B$ is the reduced state of $\omega_{AB}$.
Now, suppose that the nodes share entanglement resources, a priori, before communication begins, as illustrated in  Figure~\ref{Group_Quantum_two_nodes_network}.
Given a sufficient amount of pre-shared entanglement between the nodes,  the desired state
can be simulated if and only if the quantum communication rate satisfies   
  \begin{align}
 Q_{1,2}\geq \frac{1}{2} I(A;B)_\omega \,,
 \end{align}
  where $I(A;B)_\omega$ denotes the quantum mutual information, by the quantum  reverse Shannon theorem \cite{bennett2014quantum}. %

Coordination can be viewed as a unified framework for various models. We list a few closely-related problems:
(see Table~\ref{Table: Related problems}) 

\begin{table}[tb]
\centering
\caption{ 
Quantum Protocols 
}
\renewcommand{\arraystretch}{1}
\begin{tabular}{|>{\centering\arraybackslash}p{0.05cm}|>{\centering\arraybackslash}p{2cm}|>{\centering\arraybackslash}p{1.7cm}|>{\centering\arraybackslash}p{3.9cm}|>{\centering\arraybackslash}p{5.5cm}|>{\centering\arraybackslash}p{0.3cm}|}
\hline
 & \textbf{Task} & \textbf{Type} & \textbf{Simulated state} & \textbf{Rates} & \textbf{Ref.} \\ \hline
1 & Schumacher compression & Coordination & \(\ket{\omega_{RB}}\) &  
$Q_{1,2} \geq H(B)_{\omega}
$, 
$ R_{1,2}=0 $,
$ E_{1,2}=0 $ & \cite{schumacher1995quantum} \\ \hline
2 & Resolvability & Coordination & $\omega_{XB}$ & 
$ Q_{1,2}=0 $,
$R_{1,2}\geq I(X;B)_\omega$,
$E_{1,2}=0$& \cite{devetak2003classical} 
\\ \hline
3 & Entanglement embezzlement & Coordination & $\ket{\mu}\otimes\ket{\omega_{AB}}$ & $Q_{1,2}=R_{1,2} = 0
$ & \cite{van2003universal} \\ \hline

4 & Entanglement \mbox{dilution} & Coordination & $
\ket{\omega_{AB}}$ &  
$Q_{1,2}=0$,
$R_{1,2}\approx 0 $, 
$E_{1,2}\geq H(B)_\omega$
& \cite{harrow2004tight} \\ \hline
5 & Entanglement distillation & Distillation & $\ket{\Phi_{AB}}$ &
$Q_{1,2}=0$,
$R_{1,2}\geq H(A|B)_{\omega} $,
$E^{\text{out}}_{1,2}\leq I(A \rangle B)_{\omega} $ & \cite{
 devetak2005distillation} \\ \hline
 6 & Subspace transmission & Communication & \(\omega_{RB} = (\mathrm{id} \otimes \mathcal{N}_{A \to B})(\omega_{RA})\) & $Q^{\text{out}}_{1,2} \leq I(R \rangle B)_{\omega}\,,\;E_{1,2}=0$ & \cite{
 Devetak:05p} \\ \hline
7 & State merging 
& Coordination & $  \omega_{AB}$ & $R_{1,2}\geq I(A;B)_{\omega}\,,\;Q_{1,2}\geq H(A|B)_\omega$ & \cite{horodecki2007quantum} \\ \hline
8 & State splitting 
& 
{Coordination} &  $\ket{\omega_{ABR}}
$& $Q_{1,2}\geq \frac{1}{2}I(R;B)_\omega$, $E_{1,2}\geq \frac{1}{2}I(A;B)_\omega$ & \cite{devetak2005triangle} \\ \hline
9 & Father protocol & Communication & $\omega_{RBE} = 
(\mathrm{id}
\otimes \mathcal{U}^{\mathcal{N}}_{A \to BE})(\omega_{RA})$ & $Q_{1,2}\leq \frac{1}{2}I(R; B)_{\omega}\,,\;E_{1,2}\geq \frac{1}{2}I(R;E)_{\omega}$ & \cite{bennett2002entanglement} \\ \hline
10 & Mother protocol & Coordination & \(\ket{\Phi_{A'' B'' }}\otimes \ket{\omega_{ A B R}}\) & \( Q_{1,2}\geq \frac{1}{2}I(A;R)_\omega
\,,\;
E_{1,2}^{\text{out}}\leq \frac{1}{2}I(A;B)_\omega\) & \cite{
abeyesinghe2009mother}\\ \hline
11 & State \mbox{redistribution} & Coordination & $\ket{\omega_{ABGR}}$ & $Q_{1,2}	\geq\frac{1}{2}I(B;R|G)_{\omega}\,,\; Q_{1,2}+E_{1,2}	\geq H(B|G)_{\omega}$ & \cite{
devetak2008exact
} \\ \hline
%

%
12 & Channel simulation  
& Coordination & $\omega_{RBE} = 
(\mathrm{id}
\otimes \mathcal{U}^{\mathcal{N}}_{A \to BE})(\omega_{RA}) 
$ & $Q_{1,2} \geq \frac{1}{2}I(R;B)_{\omega} \,,\; E_{1,2}\geq \frac{1}{2}I(E;B)_{\omega}$ & \cite{bennett2014quantum}  \\ \hline

\end{tabular}
\vspace{0.1cm}

\label{Table: Related problems}
\end{table}
\begin{enumerate}
    \item 
    \textbf{Schumacher compression:} 
    In  lossless compression, Alice encodes a source $\omega_B^{\otimes n}$ and sends $nQ_{1,2}$ qubits to Bob. As Bob recovers the source, they effectively simulate a purification $\ket{\omega_{RB}}$. 
    Achievability requires 
    $
        Q_{1,2}\geq H(B)_{\omega}
    $ \cite{schumacher1995quantum}. 
    %
    \item \textbf{Channel resolvability:} 
Resolvability aims to approximate the output of a classical-quantum (c-q) channel using a uniformly distributed codebook 
\cite{hayashi2016quantum}.
The task is equivalent to the simulation of a c-q state $\omega_{XB}$ in a two-node network with a classical link. Channel resolvability can be achieved at a communication rate 
    $R_{1,2}\geq I(X;B)_{\rho}$ \cite{
    devetak2003classical,hayashi2016quantum}.
    Resolvability is also referred to as c-q soft covering \cite{strong_emprical_2018}. 
    Recently, Atif et al. \cite{atif2023quantum} have also introduced 
    fully quantum soft covering. 
    %
    \item 
    \textbf{Entanglement embezzlement:} 
    %
    Van Dam and Hayden \cite{van2003universal} showed that without any type of communication, 
    any pure bipartite entangled state
    $\ket{\omega_{AB}}$
    can be 
    simulated with ``embezzlement" from a finite catalyst \cite{jonathan1999entanglement}, i.e., while removing a small amount from the catalyst. 
    Specifically, they use 
catalyst states 
$
    \ket{\mu
    }\propto 
    \sum
    \frac{1}{\sqrt{j}}\ket{j}_A\otimes\ket{j}_B
$ of Schmidt rank $n$, and show 
that for embezzling with a fidelity $1-\varepsilon$ can be achieved with 
$n>m^{\frac{1}{\varepsilon}}$, where $m$ is the Schmidt rank of $\ket{\omega_{AB}}$. 
    %
   
    %
    \item 
    \textbf{Entanglement dilution:} Suppose that  Alice and Bob would like to prepare a joint state $\ket{\omega_{AB}}$ using local opreations and classical communication (LOCC), and a pre-shared maximally entangled state of dimension 
    $2^{nE_{1,2}}$. 
    The simulation requires $E_{1,2}\geq H(B)_\omega$ and negligible classical communication \cite{hayden2003communication,harrow2004tight}. 
    %
    \item 
    \textbf{Entanglement distillation:} Given $n$ copies of a bipartite state, $\omega_{AB}^{\otimes n}$,  Alice and Bob can distill entangled EPR pairs, $\ket{\Phi_{AB}}$, at an output rate that is bounded by the coherent information, $E_{1,2}^{\text{out}}\leq I(A\rangle B)_\omega$, 
    using 
    classical communication at a rate  $R_{1,2}\geq H(A|B)_\omega$ (see \cite{devetak2005distillation}). The analysis relies on typical subspace measurements and measurements in the Fourier transformed basis 
    \cite{bennett1996mixed,bennett1996purification,rains1999rigorous,horodecki1998mixed,vedral1998entanglement}.
     \item 
    \textbf{Subspace transmission:}
    Consider the transmission of quantum information via a quantum channel $\mathcal{N}_{A\to B}$.
    Based on the Lloyd-Shor-Devetak Theorem \cite{lloyd1997capacity,shor2002quantum,Devetak:05p}, a qubit transmission rate $Q_{1,2}$ is achievable if 
    $Q_{1,2}\leq I(R\rangle B)_{\omega}$, with respect to the output state $\omega_{RB}=(\mathrm{id} \otimes \mathcal{N}_{A \to B})(\omega_{RA})$. 
    This rate is not necessarily optimal in general. 
    \item 
    \textbf{State merging:} Consider a mixed state $\omega_{AB}$, shared between 
     Alice and Bob. 
    In the state merging protocol, 
    Alice 
    sends her part 
    to Bob 
    using classical and quantum communication, at rates $R_{1,2}\geq I(A;B)_{\omega}$ and $Q_{1,2}\geq H(A|B)_\omega$, respectively \cite{horodecki2005partial,horodecki2007quantum}.
    If $H(A| B)_\omega<0$, LOCC is sufficient and quantum communication is not required. 
    %

    %
     \item \textbf{State splitting:} 
This is the reverse task, where Alice holds both $A$ and $B$, and would like to send $B$ to Bob \cite{berta2011quantum}.
Let $\ket{\omega_{ABR}}$ be a purification.
In the state splitting protocol, Alice and Bob use quantum communication and pre-shared entanglement, at rates 
$Q_{1,2}\geq\frac{1}{2}I(R;B)_\omega$ and $E_{1,2}\geq\frac{1}{2}I(A;B)_\omega$, respectively \cite{devetak2005triangle,oppenheim2008state,devetak2004family,abeyesinghe2009mother}.
    \item 
    \textbf{The father protocol:}
Consider entanglement-assisted communication via a quantum channel $\mathcal{N}_{A\to B}$.
Given unlimited entanglement assistance, a qubit transmission rate $Q_{1,2}$ is achievable if and only if
    $Q_{1,2}\leq \frac{1}{2}I(R; B)_{\omega}$ (see \cite{bennett2002entanglement,bennett2014quantum}). 
%
   The information transmission rate above can be achieved with an entanglement rate of 
   $E_{1,2}\geq \frac{1}{2}I(R;E)_{\phi}$,  for $\omega_{RBE}=(\mathrm{id} \otimes \mathcal{U}^\mathcal{N}_{A \to B E})(\omega_{RA})$, where $\mathcal{U}^\mathcal{N}_{A \to B E}$ is a Stinespring dilation of the channel $\mathcal{N}_{A\to B}$. 
    %
    \item
    \textbf{The mother protocol:}   The protocol is known under different names, e.g., 
    quantum state
    transfer, fully quantum Slepian–Wolf (FQSW) \cite{abeyesinghe2009mother}, and coherent state merging 
    \cite{berta2016smooth,wilde2017quantum}. The mother protocol is also related to the father protocol by source-channel duality \cite{devetak2005triangle}. 
    In the course of the protocol, Alice and Bob perform state merging and also 
     distill entanglement at a rate $E^{\text{out}}_{1,2}\leq \frac{1}{2}I(A;B)_\omega$ \cite{devetak2005triangle,abeyesinghe2009mother}. 
    The main tool can be viewed as a  decoupling theorem.
    %


    \item 
    \textbf{State redistribution:}
Consider
a joint state  $\ket{\omega_{A B GR}}$, where Alice holds  $A$ and $B$, Bob has access to  $G$, and $R$ is a purifying reference system.  Alice would like to send  $B$ to Bob
\cite{devetak2008exact,Yard_Devetak_2009,luo2009channel}. 
The analysis is based on the decoupling approach as well.
The state redistribution setting was also considered in the one-shot case by Berta~et~al.~\cite{berta2016smooth}.
State redistribution generalizes several protocols, such as 
Schumacher's compression \cite{schumacher1995quantum}, state merging \cite{horodecki2005partial,horodecki2007quantum} 
and  
splitting \cite{abeyesinghe2009mother}.
   \item 
    \textbf{Channel simulation:} 
    According to
    the classical reverse Shannon theorem \cite{bennett2002entanglement}, a classical channel of capacity $C$ can be simulated at a classical rate $R_{1,2}$ if and only if $R_{1,2}\geq C$, given sufficient 
 common randomness \cite{cuff2008communication}. 
    The quantum analog is not  necessarily true in general, yet it holds  for a product input state, 
        $\omega_{RA}^{\otimes n}$     \cite{bennett2014quantum}.
    In this case, achievability was shown for 
    $
        Q_{1,2}\geq\frac{1}{2}I(R;B)_{\omega}
        $ and 
         $E_{1,2}\geq \frac{1}{2}I(B;E)_{\omega}
    $, for $\omega_{RBE}=(\mathrm{id} \otimes \mathcal{U}^\mathcal{N}_{A \to B E})(\omega_{RA})$. 
    The analysis shows that given pre-shared 
    entanglement embezzlement, channel simulation can be achieved without 
    backward communication \cite{bennett2014quantum}. Given LOCC, 
    the 
    entanglement cost for 
    simulation is related to 
    the entanglement of formation
    \cite{berta2013entanglement}.
\end{enumerate}

Multi-user versions of the protocols above have also been studied extensively in recent years. 
 Abeyesinghe et. al. \cite{abeyesinghe2009mother}   use the mother protocol to generate  distributed compression protocols for correlated quantum sources. Other results on quantum distributed compression can be found in \cite{ahn2006distributed,KhanianWinter:18a,salek2018quantum,Faithful_simulation_Heidari_2019}.
Simulation of broadcast and multiple-access channels is considered in \cite{cheng2023quantum,cao2024channel} and 
 \cite{nema2024one}, respectively. 
  George and Cheng~\cite{george2024coherent} have recently studied multipartite state splitting. 
    Multi-user distillation and manipulation is considered in  \cite{smolin2005entanglement,bravyi2006ghz,augusiak2009multipartite,streltsov2017rates,murta2020quantum,Salek_2022_Winter,salek2023new}.
 Streltsov et al. \cite{streltsov2020rates} studied 
 multipartite state merging.  
    A father protocol for broadcast and multiple-access channels is presented in 
    \cite{dupuis2010father} and \cite{hsieh2008entanglement}, respectively.  

     This brief overview is by
no means exhaustive and is only meant 
to provide some
background in order to place this contribution in context.

\begin{figure}[t]
    \centering
    \begin{minipage}[t]{0.3\textwidth} 
        \centering
        \includegraphics[scale=0.68,trim={4cm -1cm 4cm -0.5cm}]{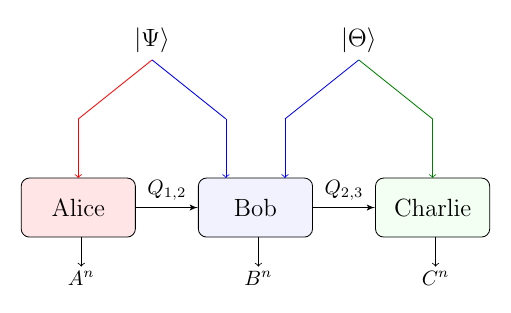}
        \caption{Cascade network  with quantum links and pre-shared entanglement}
        \label{Group_Cascade_network}
    \end{minipage}
    \hfill
    \begin{minipage}[t]{0.3\textwidth} 
        \centering
        \includegraphics[scale=0.68,trim={7cm 0 6cm -0.5cm}]{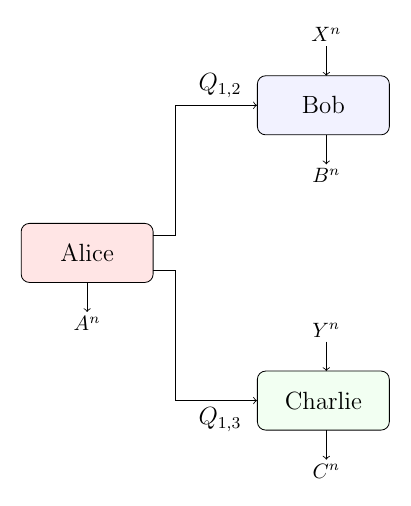}
        \caption{Broadcast network  with quantum links}
        \label{Group_Broadcast_network}
    \end{minipage}
    \hfill
    \begin{minipage}[t]{0.3\textwidth} 
        \centering
        \includegraphics[scale=0.68,trim={6cm 0 6cm -0.5cm}]{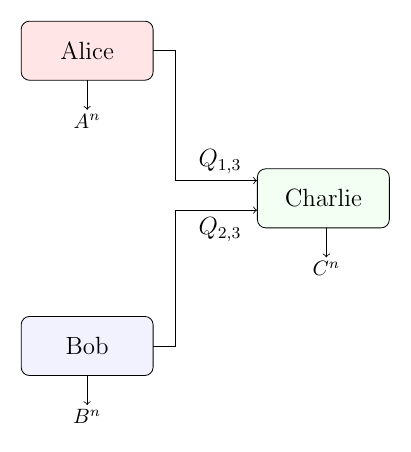}
        \caption{Multiple-access network  with quantum links}
        \label{Group_Multiple_access_network}
    \end{minipage}
    \label{fig:full2}
\end{figure}

\paragraph*{Multi-node quantum coordination}
 In this work, we consider   quantum coordination in   networks with  quantum links. 
 Coordination in multi-user networks with quantum links is motivated by applications  
 such as the quantum Internet and quantum repeaters \cite{6253218}. In each network, we determine the optimal coordination rates,  
characterizing the minimal resources required in order to simulate a joint quantum state among multiple parties. 
 We further discuss the implications of our results on  nonlocal quantum games. In particular,   coordination in the broadcast network in Figure~\ref{Group_Broadcast_network} can be viewed as a sequential game, where  a coordinator (the sender)  provides the players (the receivers) 
 with quantum resources.
 In the course of the game, the referee sends questions, $X^n$ and $Y^n$, to 
 each player,
 and they respond with $B^n$ and $C^n$. In order to win the game with a certain probability, the communication rates must satisfy the constraints 
 with respect to an appropriate correlation.

%

 First, we examine a cascade network that consists of three users, Alice, Bob, and Charlie, as depicted in Figure~\ref{Group_Cascade_network}. 
 The cascade  network is of particular importance for quantum repeater systems, which are crucial for  the development of the quantum Internet \cite{azuma2023quantum,PeregDeppeBoche:21p2} and long-range quantum key distribution (QKD) \cite{jiang2009quantum,langenfeld2021quantum}. In this network,
 Alice, Bob, and Charlie wish to simulate a joint quantum state $\omega_{ABC}$. Let $\ket{\omega_{ABCR}}$ be a purification of the desired state.
 Before communication begins, each party shares entanglement with their nearest neighbor,  at a limited rate.
Now, Alice sends qubits to Bob at a rate $Q_{1,2}$, and thereafter,  Bob sends qubits to Charlie at a rate $Q_{2,3}$.
We show that coordination can be achieved if and only if the quantum communication rates $Q_{1,2}$ and $Q_{2,3}$ satisfy
\begin{align}
Q_{1,2}        &\geq \frac{1}{2} I(BC;R)_\omega\,,\\
Q_{1,2}+E_{1,2}&\geq H(BC)_{\omega}\,, \\
Q_{2,3}  &\geq \frac{1}{2} I(C;RA)_\omega\,,\\
Q_{2,3}+E_{2,3} &\geq H(C)_{\omega}
\,,
\end{align}
where $E_{i,j}$ is the entanglement rate between Node $i$ and Node $j$.

Next, we consider a broadcast network with one sender, Alice, and two receivers, Bob and Charlie, where the receivers are provided with classical sequences of information $X^n$ and $Y^n$. See Figure~\ref{Group_Broadcast_network}.
Among others, this is motivated by applications that are based on entanglement distribution \cite{bauml2017fundamental}.
We show that coordination can be achieved if and only if the quantum communication rates $Q_{1,2}$ and $Q_{1,3}$ satisfy
 \begin{align}
 Q_{1,2}&\geq H(B|X)_{\omega} \,,
 \label{Equation:Broadcast_Q12}
 \\
 Q_{1,3}&\geq H(C|Y)_{\omega} \,.
 \label{Equation:Broadcast_Q13}
 \end{align}
 In the third setting, we consider a  multiple-access network, with two transmitters, Alice and Bob, and one receiver, Charlie, as illustrated in Figure~\ref{Group_Multiple_access_network}. This network is relevant for  multiple access QKD \cite{6253218,zhang2020design} and IoT \cite{9232550}.  %

We observe that since there is no cooperation between the transmitters, a joint state $\omega_{ABC}$ can only be simulated if it is isometrically equivalent to a state of the form
 $\phi_{AC_1}\otimes \chi_{BC_2}$. %
 Then, we show that
coordination can be achieved if and only if
 \begin{align}
     Q_{1,3}&\geq H(C_1)_{\phi} \,,
     \\
      Q_{2,3}&\geq H(C_2)_{\chi} \,.
     \end{align}

 Furthermore, we discuss the implications of our results on  nonlocal quantum games. In particular,   coordination in the broadcast network in Figure~\ref{Group_Broadcast_network} can be viewed as a sequential game, where Alice is a coordinator that provides the players, Bob and Charlie, with quantum resources.
 In the course of the game, the referee sends questions, $X^n$ and $Y^n$, to Bob and Charlie respectively, and they send their response through $B^n$ and $C^n$. In order to win the game with a certain probability, the communication rates must satisfy the constraints \eqref{Equation:Broadcast_Q12}-\eqref{Equation:Broadcast_Q13}
 with respect to an appropriate correlation.

In the analysis, we use different techniques for each setting. For the cascade network, we use the state redistribution theorem \cite{Yard_Devetak_2009}. In the broadcast network, we assume that Alice does not have prior correlation with Bob and Charlie's resources $X^n$ and $Y^n$. Therefore, the standard techniques of state redistribution \cite{Yard_Devetak_2009} or quantum source coding with side information \cite{9039682} are not suitable for our purposes. Instead, we use
a quantum version of binning. In the analysis of the multiple-access network, we use the Schumacher compression protocol and the isometric relation that is dictated by the network topology. 

The paper is organized as follows. In 
Section~\ref{Section:Definitions_and_results}, we define three coordination models and present our main results for each  network. In 
Section~\ref{Section:Nonlocal_games},
we discuss the implications of our results on nonlocal games, and in Sections \ref{Cascade_network_analysis}, \ref{Broadcast_network_analysis}, and \ref{Multiple_access_network_analysis}, we provide the analysis for the cascade, broadcast, and multiple-access networks,  respectively.

\section{Model Definitions and Results}
\label{Section:Definitions_and_results}
We introduce three quantum coordination models and provide the required definitions. %

The following notation conventions are used throughout. 
We use uppercase letters
$X,Y,Z,\ldots$ for discrete random variables on finite alphabets   $\mathcal{X},\mathcal{Y},\mathcal{Z},...$,  and lowercase 
$x,y,z,\ldots$ for their realization, respectively.
Let $x^n=(x_i)_{i\in [n]}$ represent  
a sequence of letters from  
 $\mathcal{X}$. 
 A quantum state is specified by 
a density operator, $\rho_A$, on the Hilbert space $\mathcal{H}_A$.
Let $\Delta(\mathcal{H}_A)$ denote the set of all density 
operators on $\mathcal{H}_A$. Then, $A^n=A_1\cdots A_n$ is  
a sequence of quantum systems whose joint state $\rho_{A^n}$ belongs to $\Delta(\mathcal{H}_A^{\otimes n})$.
A quantum  channel is a completely-positive trace-preserving map $\mathcal{E}_{A\to B}:\Delta(\mathcal{H}_A)\to \Delta(\mathcal{H}_B)$.
The quantum mutual information is defined as
$%
I(A;B)_\rho=H(\rho_A)+H(\rho_B)-H(\rho_{AB}) 
$, %
where $H(\rho) \equiv -\trace[ \rho\log(\rho) ]$ is the von Neumann entropy.
The conditional quantum entropy is defined as 
$H(A|B)_{\rho}=H(\rho_{AB})-H(\rho_B)$,  and
the conditional quantum mutual information as
$I(A;B|C)_{\rho}=H(A|C)_\rho+H(B|C)_\rho-H(AB|C)_\rho$.

\subsection{Cascade network}
\label{Cascade_subsection}
\begin{figure}[bt]
\center
\includegraphics[scale=0.8,trim={5.3cm 0 5.5cm 0}]
{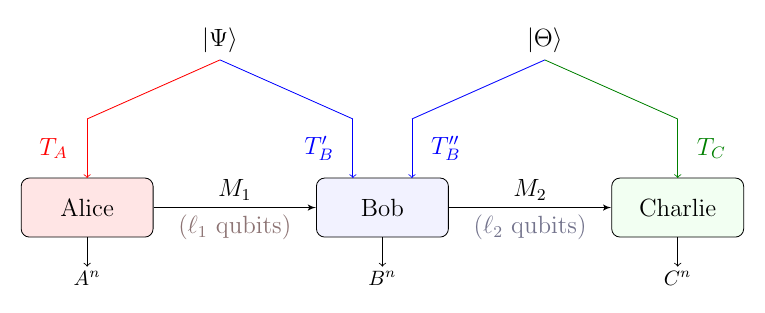} %
\caption{Cascade network with rate-limited entanglement. 
  Before communication begins, each party shares bipartite entanglement with their nearest neighbor. 
%
 Alice prepares the state of her output  $A^{n}$, as well as a ``quantum description"  $M_{1}$ consisting of $\ell_1$ qubits. She sends $M_{1}$ to Bob. As Bob receives $M_{1}$, he encodes the output $B^{n}$, along with his own quantum description, $M_{2}$ consisting of $\ell_2$ qubits. Next, Bob  sends $M_{2}$ to Charlie. Upon receiving $M_{2}$, Charlie prepares the output state for $C^{n}$. 
The objective is to simulate $n$ copies of a desired quantum state $\ket{\omega_{RABC}}$, i.e., %
for the encoded state  $\widehat{\rho}_{R^{n}A^{n}B^{n}C^n}$ to be arbitrarily close to  $\ket{\omega_{RABC}}^{\otimes n}$.}
\label{Cascade_network}
\end{figure}

Consider the cascade network with rate-limited entanglement, as depicted in Figure~\ref{Cascade_network}.  In the Introduction section, we used the notation
$Q_{i,j}$ for the communication rate from Node $i$ to Node $j$.
Here, 
we simplify the notation, and write $Q_1\equiv Q_{1,2}$ and
$Q_2\equiv Q_{2,3}$, for convenience.

Alice, Bob, and Charlie would like to simulate a joint state $\omega_{ABC}^{\otimes n}$, where $\omega_{ABC}\in\Delta(\mathcal{H}_{A}\otimes\mathcal{H}_{B}\otimes\mathcal{H}_{C})$.
Let $\ket{\omega_{RABC}}$ be a purification of $\omega_{ABC}$, where $R$ can be viewed as Alice's reference.
Before communication begins, each party shares bipartite entanglement with their nearest neighbor. 
The bipartite state $\ket{\Psi_{T_{A}T_{B}'}}$ indicates
the entanglement resource shared between Alice and Bob, while $\ket{\Theta_{T_{B}'' T_{C}}}$ is shared between Bob and Charlie.  
The coordination protocol begins with Alice preparing the state of her output system $A^{n}$, as well as a ``quantum description"  $M_{1}$. She sends $M_{1}$ to Bob. As Bob receives $M_{1}$, he encodes the output $B^{n}$, along with his own quantum description, $M_{2}$. Next, Bob  sends $M_{2}$ to Charlie. Upon receiving $M_{2}$, Charlie prepares the output state for $C^{n}$. 

The transmissions $M_{1}$ and $M_{2}$ are limited to the quantum communication rates $Q_{1}$ and $Q_{2}$, while the pre-shared resources between Alice and Bob
and between Bob and Charlie are limited to the entanglement rates $E_1$ and $E_2$, respectively.

\begin{definition}
\normalfont
A 
$(2^{\ell_{1}},2^{\ell_{2}},2^{k_1},2^{k_2},n)$ coordination code for the cascade network %
in Figure~\ref{Cascade_network} consists of: 
\begin{enumerate}[]
\item
Two bipartite states 
$\ket{\Psi_{T_A T_B' }}$ and
$\ket{\Theta_{T_B'' T_C }}$ on
Hilbert spaces  of dimension
$2^{k_1}$ and $2^{k_2}$, respectively, i.e.,
\begin{align} 
\mathrm{dim}(\mathcal{H}_{T_A})&=
\mathrm{dim}(\mathcal{H}_{T_B'})=2^{k_1}
\,,
\\
\mathrm{dim}(\mathcal{H}_{T_B''})&=
\mathrm{dim}(\mathcal{H}_{T_C})=2^{k_2}
\,, \end{align}
\item
two Hilbert spaces, $\mathcal{H}_{M_{1}}$ and $\mathcal{H}_{M_{2}}$, of dimension
\begin{align} \mathrm{dim}(\mathcal{H}_{M_{j}})=2^{\ell_{j}}\;\text{ for \ensuremath{j\in\{1,2\}}} \,, \end{align} 
and 
\item
three encoding maps, 
\begin{align} \mathcal{E}_{ \bar{A}^n T_A\to A^{n}M_{1}}&	:\Delta(\mathcal{H}_{A}^{\otimes n}\otimes\mathcal{H}_{T_A})\to\Delta(\mathcal{H}_{A}^{\otimes n}\otimes\mathcal{H}_{M_{1}})\,,
\\
\mathcal{F}_{M_{1}T_B' T_B''\to B^{n}M_{2}}&	:\Delta(\mathcal{H}_{M_{1}}\otimes\mathcal{H}_{T_B'T_B''})
\to
\Delta(\mathcal{H}_{B}^{\otimes n}\otimes\mathcal{H}_{M_{2}})
\,,
\intertext{and}	
\mathcal{D}_{M_{2}T_C\to C^{n} } &	:\Delta(\mathcal{H}_{M_{2}}\otimes\mathcal{H}_{T_C})
\to
\Delta(\mathcal{H}_{C}^{\otimes n})\,,
\end{align}
corresponding to Alice, Bob, and Charlie, respectively. 
\end{enumerate}
\end{definition}
The coordination protocol has limited communication rates $Q_j$ and entanglement rates $E_j$, for $j\in \{1,2\}$. 
That is, before the protocol begins, Alice and Bob are provided with $k_1=nE_1$ qubit pairs, while Bob and Charlie share $k_2=nE_2$  pairs. 
During the protocol, Alice transmits $\ell_1=nQ_1$ qubits to Bob, and then Bob transmits $\ell_2=nQ_2$ qubits to Charlie. See Figure~\ref{Cascade_network}.
Hence, $E_j=\frac{k_j}{n}$ is the rate of entanglement pairs per output, and $Q_j=\frac{\ell_j}{n}$ is the rate of qubit transmissions per output, for $j\in \{1,2\}$.
A  detailed description of the protocol is given below.

The coordination protocol works as follows.
 Alice prepares the state $\omega_{R\bar{A}}^{\otimes n}$ locally, and
 applies the encoding map $ \mathcal{E}_{\bar{A}^n T_A\to A^{n}M_{1}}$ on her share $T_A$ of the entanglement resources.
This results in the output state
\begin{align} \rho_{R^{n}A^{n}M_{1}T_B'}^{(1)}=(\mathrm{id}_{R^n}\otimes\mathcal{E}_{\bar{A}^n T_A\to A^{n}M_{1}}\otimes \mathrm{id}_{T_B'})(\omega_{R\bar{A}}^{\otimes n}\otimes \Psi_{T_A T_B'})
\,. 
\label{Cascade_rho_1}
\end{align}
She sends $M_{1}$ to Bob. Having received  $M_{1}$, Bob uses it along with his share 
$T_B' T_B''$ of the entanglement resources to encode the systems $B^{n}$ and $M_{2}$. To this end, he uses the map $\mathcal{F}_{M_{1}T_B' T_B'' \to B^{n}M_{2}}$, hence 
\begin{align}
\rho_{R^{n}A^{n}B^{n}M_{2}T_C}^{(2)}=(\mathrm{id}_{R^{n}A^{n}}\otimes\mathcal{F}_{M_{1}T_B' T_B''\to B^{n}M_{2}}\otimes 
\mathrm{id}_{T_C})
(\rho_{R^{n}A^{n}M_{1} T_B'}^{(1)}
\otimes
\Theta_{T_B''\, T_C})\,.
\label{Cascade_rho_2}
\end{align}
Bob sends $M_{2}$ to Charlie, who applies the encoding channel $\mathcal{D}_{M_{2}T_C\to C^{n}}$. This results in the final joint state, 
\begin{align} 
\widehat{\rho}_{R^{n}A^{n}B^{n}C^{n}}=(\mathrm{id}_{R^{n}A^{n}B^{n}}\otimes\mathcal{D}_{M_{2}T_C\to C^{n}})\left(\rho_{R^{n}A^{n}B^{n}M_{2}T_C}^{(2)}\right)\,.
\label{Cascade_rho_hat}
\end{align}
The objective %
is that the final state  $\widehat{\rho}_{R^n A^{n}B^{n}C^{n}}$ is arbitrarily close to the desired state $\omega_{RABC}^{\otimes n}$.
\begin{definition} 
\normalfont
A rate tuple $(Q_{1},Q_{2},E_1,E_2)$ is achievable, if for every $\varepsilon,\delta>0$ and sufficiently large $n$, there exists a $(2^{n(Q_{1}+\delta)},2^{n(Q_{2}+\delta)},2^{n(E_{1}+\delta)},2^{n(E_{2}+\delta)},n)$ coordination code satisfying 
\begin{align}
\norm{\widehat{\rho}_{R^n A^{n}B^{n}C^{n}}-\omega_{RABC}^{\otimes n}}_{1}\leq\varepsilon.
\end{align} 
\end{definition}
\begin{remark}
\label{Remark:Resource_Cascade}
Coordination in the cascade network can also be represented as 
as a resource inequality \cite{DevetakHarrowWinter:08p}
\begin{align}
Q_1 [q\to q]_{A\to B}+E_1 [q q]_{A B}+Q_2 [q\to q]_{B\to C}
+E_2 [q q]_{BC} \geq \left\langle \omega_{ABC} \right\rangle
\end{align}
where the resource units $[q\to q]$, $[q q]$, and $\left\langle \omega_{ABC} \right\rangle$ represent a single use of a noiseless qubit channel, an EPR pair, and the desired state $\omega_{ABC}$, respectively. 
\end{remark}
The optimal coordination rates for the cascade network are established below. %
\begin{theorem}
\normalfont
\label{Theorem_Cascade}
Consider a desired state $\ket{\omega_{RABC}}$.
A rate tuple
$\left(Q_{1},Q_{2},E_{1},E_{2}\right)$ is achievable for
     coordination in the cascade network %
     in Figure~\ref{Cascade_network}, if and only if
\begin{align}
Q_{1}        &\geq \frac{1}{2}I(BC;R)_\omega\,,\\
Q_{1}+E_{1}&\geq H(BC)_{\omega}\,, \\
Q_{2}  &\geq \frac{1}{2}I(C;RA)_\omega\,,\\
Q_{2}+E_{2} &\geq H(C)_{\omega}
\,.
\end{align}
\end{theorem}
The proof for Theorem~\ref{Theorem_Cascade} is provided in 
Section~\ref{Cascade_network_analysis}.

\begin{corollary}
\normalfont
\label{Theorem_Cascade_Pure}
For a pure state $\ket{\omega_{ABC}}$, 
 the coordination capacity region for the cascade network is given by the set
%
\begin{align}
\mathcal{Q}_{\text{Cascade}}(\omega)=\left\{ 
\begin{array}{rrl}
(Q_{1},E_{1},Q_{2},E_{2}):
& Q_{1}+E_{1} &\geq H(BC)_{\omega}\,,
\\ 
& Q_{2}  &\geq \frac{1}{2} I(C;A)_\omega\,,\;
\\ 
&Q_{2}+E_{2} &\geq H(C)_{\omega}
\,
\end{array}
\right\} \,.
\end{align}
\end{corollary}

%
%
The examples below demonstrate that coordination of entanglement and coordination of separable correlations behave differently. 

\begin{figure}[t]
    \centering
    \begin{minipage}[b]{0.4\linewidth} 
        \centering
        \includegraphics[scale=0.56]{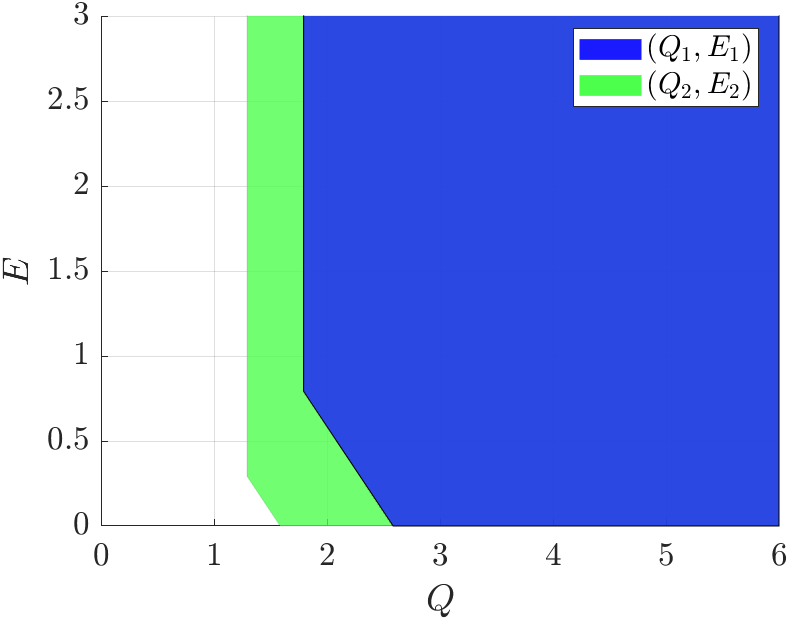} 
        \vspace{-0.5cm}
        \begin{center}
        \footnotesize{(a) Example~\ref{Mixed_example}: Mixture} 
        \end{center}
    \end{minipage}
    \hspace{2.5cm}
    \begin{minipage}[b]{0.4\linewidth} 
        \centering
        \includegraphics[scale=0.56]{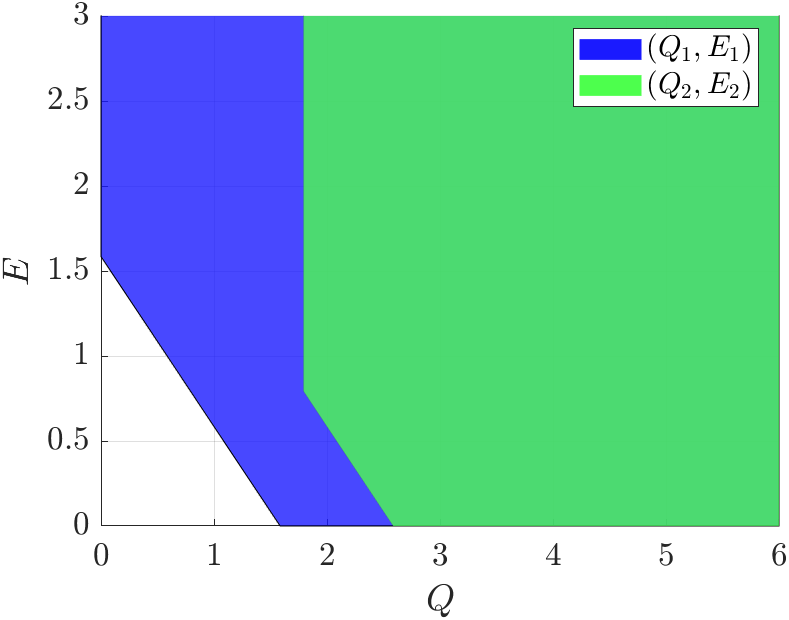} 
        \vspace{-0.5cm}
        \begin{center}
        \footnotesize{(b) Example~\ref{Pure_example}: Entanglement}
        \end{center}
    \end{minipage}
    \caption{Achievable coordination regions in two examples. In the mixed-state example, Alice is required to send qubits to Bob at a \emph{higher} rate than  Bob  to Charlie. 
    In the entangled-state example,
    Alice is required to send qubits to Bob at a  rate  that is \emph{lower} than  Bob  to Charlie. This occurs because of the ``knowing less than nothing" phenomenon, i.e., the entropy of a subsystem is larger than the joint entropy.  
    }
    \label{Figure_coordination_capacity_cascade}
\end{figure}

\begin{example}
[Mixture]
\label{Mixed_example}

Let $\mathcal{H}_A$, $\mathcal{H}_B$, and $\mathcal{H}_C$ be Hilbert spaces of dimension $3$, i.e., qutrits. 
Consider the simulation of a mixed state, 
\begin{align}
\hspace{-0.2cm} \omega_{ABC}=\frac{1}{6}\left(\ketbra{012}+\ketbra{021}+\ketbra{102}+\ketbra{120}+\ketbra{201}+\ketbra{210}\right)
\end{align}
The example is analogous to classical task assignment \cite[Example 3]{cuff2010coordination}.
The state above is thus purified by
\begin{align}
    \ket{\omega_{RABC}} 
    &= \frac{1}{6} \big( 
    \ket{0} \otimes \ket{012} 
    + \ket{1}\otimes\ket{021}  
    + \ket{2} \otimes \ket{102} 
    + \ket{3}\otimes\ket{120}  
    + \ket{4}\otimes\ket{201}  
    + \ket{5}\otimes\ket{210}  \big)
\end{align}
where $\{\ket{i}\}_{i=0,\ldots,5.}$ forms an orthonormal basis for the reference system $R$.
%
In this case,  coordination can be achieved if and only if
the  rate tuple $(Q_{1},E_{1},Q_{2},E_{2})$ belongs to the following set,
\begin{align}
\left\{ 
\begin{array}{rrl}
(Q_{1},E_{1},Q_{2},E_{2}):
&Q_{1} & \geq 
1.7925, \\
&Q_{1} + E_{1} & \geq 
2.5850, \\
&Q_{2} & \geq 
1.2925, \\
&Q_{2} + E_{2} & \geq 
1.5850.
\end{array}
\right\} \,.
\end{align}
The coordination capacity region $\mathcal{Q}_{\text{Cascade}}(\omega)$ is illustrated in 
Figure~\ref{Figure_coordination_capacity_cascade} (a), where the blue region shows the tradeoff between Alice's rates, $Q_1$ and $E_1$, and the green region is associated with Bob's rates, $Q_2$ and $E_2$.

Suppose that $E_1=E_2$.
As can be seen in the figure,  Alice is required to send qubits to Bob at a higher rate than  Bob  to Charlie. This is intuitive since Alice  encodes  information for both Bob and Charlie, whereas Bob is only encoding Charlie's  information.

\end{example}

\begin{example}[Entanglement]
\label{Pure_example}

Consider the simulation of a pure tripartite entangled state, 
\begin{align}
\ket{\psi_{ABC}}=\frac{1}{\sqrt{6}}\left(\ket{012}+\ket{021}+\ket{102}+\ket{120}+\ket{201}+\ket{210}\right)
\end{align}
According to Corollary~\ref{Theorem_Cascade_Pure},
$\ket{\psi_{ABC}}^{\otimes n}$ can be simulated if and only if
the  rate tuple $(Q_{1},E_{1},Q_{2},E_{2})$ belongs to the following set,
\begin{align}
\left\{ 
\begin{array}{rrl}
(Q_{1},E_{1},Q_{2},E_{2}):
&Q_{1}+E_{1}&\geq 
1.5850\,, \nonumber\\
&Q_{2}  &\geq 
0.7925\,,\nonumber\\
&Q_{2}+E_{2} &\geq 
1.5850
\,.\nonumber
\end{array}
\right\} \,.
\end{align}
The coordination capacity region $\mathcal{Q}_{\text{Cascade}}(\psi)$ is illustrated in 
Figure~\ref{Figure_coordination_capacity_cascade} (b). As before,  the blue region shows the tradeoff between Alice's rates, $Q_1$ and $E_1$, and the green region is associated with Bob's rates, $Q_2$ and $E_2$.

Suppose that $E_1=E_2$.
Here, as opposed to Example~\ref{Mixed_example},   Alice is required to send qubits to Bob at a  rate  that is \emph{lower} than  Bob  to Charlie. This occurs because of the ``knowing less than nothing" phenomenon \cite{GourWildeBrandsenGeng:22a}. That is, in the presence of entanglement,  a subsystem can have a larger entropy compared to   the joint system. 
The behavior in each example is completely different. 
\end{example}

\subsection{Broadcast network}
\label{Broadcast_subsection}
\begin{figure}[bt]
\center
\includegraphics[scale=0.8,trim={5.3cm 0 5.5cm 0}]
{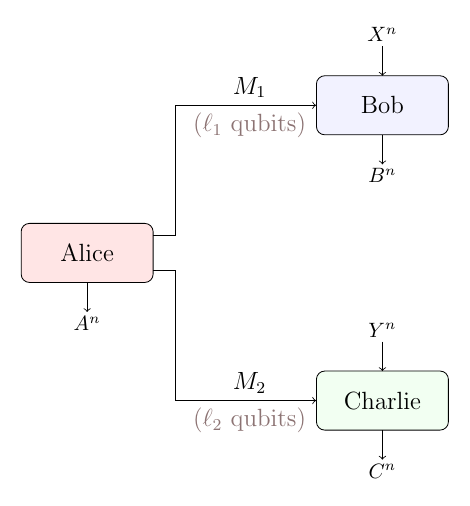} %
\caption{Broadcast network.   
%
Before communication takes place,
the classical sequences $X^n$ and $Y^n$ are drawn from a common source, and given to Bob and Charlie, respectively.
 Alice prepares the state of her output  $A^{n}$, as well as her ``quantum descriptions",  $M_{1}$ and $M_2$ consisting of $\ell_1$ and $\ell_2$ qubits respectively. She transmits $M_{1}$ and $M_2$ to Bob and Charlie, respectively. As Bob receives $M_{1}$, he encodes the output $B^{n}$. Similarly, Charlie receives   $M_{2}$, and encodes $C^{n}$. 
The objective is to simulate $n$ copies of a desired quantum state $\omega_{ABC}$, i.e., %
for the encoded state  $\widehat{\rho}_{A^{n}B^{n}C^n}$ to be arbitrarily close to  $\omega_{ABC}^{\otimes n}$.}
\label{Broadcast_network}
\end{figure}

Consider the broadcast network in Figure~\ref{Broadcast_network}. This network, can be useful in analyzing refereed games and the required resources for achieving certain performances as described in section \ref{Section:Nonlocal_games}.
 As before, we simplify the notation
$Q_{i,j}$ from the Introduction section, and write $Q_1\equiv Q_{1,2}$ and
$Q_2\equiv Q_{1,3}$, for convenience.
Consider a classical-quantum state,
\begin{align}
\omega_{XYABC}=\sum_{x\in\mathcal{X}} \sum_{y\in\mathcal{Y}}
p_{XY}(x,y)\ketbra{x,y}_{X,Y}
\otimes \ketbra{\sigma^{(x,y)}_{ABC}}
\label{Equation:Broadcast_omega_XYABC}
\end{align}
corresponding to a given ensemble of states $\left\{p_{XY},\ket{\sigma^{(x,y)}_{ABC}}\right\}$ in $\Delta(\mathcal{H}_A\otimes\mathcal{H}_B\otimes\mathcal{H}_C)$.

Alice, Bob, and Charlie would like to simulate  $\omega_{XYABC}$.
Before communication takes place,
the classical sequences $X^n$ and $Y^n$ are drawn from a common source $p_{XY}^{\otimes n}$.
The sequence $X^n$ is given to Bob, while $Y^n$ is given to Charlie (see Figure~\ref{Broadcast_network}).

Initially, Alice encodes her output $A^{n}$, along with two quantum descriptions, $M_{1}$ and $M_{2}$. She then transmits   $M_{1}$ and $M_2$, to Bob and Charlie, respectively,  at limited qubit transmission rates, $Q_1$ and $Q_2$.
As Bob receives the quantum description $M_{1}$, he uses it together with the classical sequence  $X^n$ to   encode the output $B^{n}$. Similarly, Charlie receives $M_{2}$ and  $Y^n$, and encodes his output $C^{n}$. %
\begin{definition} 
\normalfont
A $(2^{\ell_1},2^{\ell_2},n)$ coordination code for the broadcast network with side information described in Figure~\ref{Broadcast_network}, consists of
two Hilbert spaces, $\mathcal{H}_{M_{1}}$ and $\mathcal{H}_{M_{2}}$, of dimensions
\begin{align} 
\mathrm{dim}(\mathcal{H}_{M_{j}})=2^{\ell_{j}}\;\text{for \ensuremath{j\in\{1,2\}} \,,} 
\end{align} 
and three encoding maps,
\begin{align}
\mathcal{E}_{A^{n}\to A^{n}M_{1}M_{2}} &: \Delta(\mathcal{H}_{A}^{\otimes n}) \to \Delta(\mathcal{H}_{A}^{\otimes n} \otimes \mathcal{H}_{M_{1}} \otimes \mathcal{H}_{M_{2}}),
\\
\mathcal{F}_{X^n M_{1}\to B^{n}} &: \mathcal{X}^n\otimes  \Delta(\mathcal{H}_{  M_{1}}) \to \Delta(\mathcal{H}_{ B}^{\otimes n}),
\intertext{and}
\mathcal{D}_{Y^nM_{2}\to  C^{n}} &: \mathcal{Y}^n\otimes \Delta(\mathcal{H}_{ M_{2}}) \to \Delta(\mathcal{H}_{C}^{\otimes n}).
\end{align}
\end{definition}
corresponding to Alice, Bob, and Charlie, respectively.
In the course of the protocol, Alice transmits $\ell_1=nQ_1$ qubits to Bob, and $\ell_2=nQ_2$ qubits to Charlie, as illustrated in Figure~\ref{Broadcast_network}. Thus, the   qubit transmission rates are $Q_j=\frac{\ell_j}{n}$ for $j\in \{1,2\}$. %

\begin{remark}
In the quantum world, broadcasting a quantum state among multiple receivers is impossible by the no-cloning theorem. However, in the broadcast network in Figure~\ref{Broadcast_network}, Alice  sends two different ``quantum messages" $M_1$ and $M_2$ to Bob and Charlie, respectively. %
Roughly speaking, Alice is broadcasting correlation. 
Since Alice prepares both quantum descriptions, $M_1$ and $M_2$, she can create correlation and generate tripartite entanglement between her, Bob, and Charlie.
\end{remark}

The  coordination protocol is described below. 
Alice applies her encoding map %
and prepares %
\begin{align} 
\rho_{A^{n}M_{1}M_{2}}^{(1)}=
 \mathcal{E}_{A^{n}\to A^{n}M_{1}M_{2}}(\omega_{A}^{\otimes n}) \,. 
\end{align}
She sends  $M_{1}$ and $M_{2}$ to Bob and Charlie, respectively. Once Bob receives $M_{1}$ and the classical assistance, $X^n$, he applies his encoding map 
$\mathcal{F}_{X^n M_{1}\to B^{n}}$. 
Similarly,  Charlie receives $M_2$ and $Y^n$, and applies $\mathcal{D}_{Y^nM_{2}\to  C^{n}}$. Their encoding operations result in the following extended state:
\begin{multline}
\widehat{\rho}_{X^n Y^n A^{n}B^{n}C^{n}}
=
\sum_{x^n\in\mathcal{X}^n} \sum_{y^n\in\mathcal{Y}^n}
 p_{XY}^{\otimes n}(x^n,y^n) \ketbra{x^n,y^n}_{X^n Y^n}\otimes
 \\
 (\mathrm{id}_{A^{n}}\otimes\mathcal{F}_{X^{n}M_{1}\to B^{n}}\otimes\mathcal{D}_{Y^{n}M_{2}\to C^{n}})
 \left(
 \ketbra{x^n,y^n}_{\bar{X}^n \bar{Y}^n} \otimes\rho_{A^{n}M_{1}M_{2}}^{(1)}
 \right)
\,,
\end{multline}
where $\bar{X}^n \bar{Y}^n$ are classical registers that store a copy of the (classical) sequences $X^n Y^n$, respectively.
The goal is to encode such that  the final state  $\widehat{\rho}_{X^n Y^n A^{n}B^{n}C^{n}}$ is arbitrarily close to the desired state $\omega_{XYABC}^{\otimes n}$.
\begin{definition}
\normalfont
A rate pair $(Q_{1},Q_{2})$ is achievable, if for every $\varepsilon,\delta>0$ and sufficiently large $n$, there exists a $(2^{n(Q_{1}+\delta)},2^{n(Q_{2}+\delta)},n)$ coordination code satisfying 
\begin{align}
\norm{\widehat{\rho}_{X^n Y^n A^{n}B^{n}C^{n}}-\omega_{XY ABC}^{\otimes n}}_{1}\leq\varepsilon.
\end{align} 
\end{definition}
 \begin{remark}
 \label{Remark:Broadcast_no_correlation}
Notice that Alice has no access to $X^n$ nor $Y^n$.
Therefore, coordination can only be achieved for states 
$\omega_{XYABC}$ such that there is no correlation between 
$A$ and $XY$, on their own. That is, the reduced state $\omega_{XYA}$ must have a product form, 
\begin{align}
\omega_{XYA}=\omega_{XY}\otimes \omega_A \,.
\label{Equation:Broadcast_XYA}
\end{align}
Since Alice does not share prior correlation with Bob and Charlie's resources $X^n$ 
and $Y^n$,  standard techniques, such as state redistribution \cite{Yard_Devetak_2009} and quantum source coding with side information \cite{9039682}, are not suitable for our purposes. Instead, we introduce
a quantum version of binning. 
 \end{remark}
 
The optimal coordination rates for the broadcast network are established below. 
\begin{theorem}
\normalfont
\label{Theorem_Broadcast}
    A rate pair $(Q_1,Q_2)$ for the broadcast network in Figure~\ref{Broadcast_network} is achievable if and only if
\begin{align}
Q_{1}        &\geq H(B|X)_\omega\,,\\
Q_{2}  &\geq H(C|Y)_\omega
\end{align}
\end{theorem}
The proof for 
Theorem~\ref{Theorem_Broadcast} is provided in Section~\ref{Broadcast_network_analysis}.

\subsection{Multiple access network}
\label{MAC_subsection}
\begin{figure}[bt]
\center
\includegraphics[scale=0.8,trim={5.3cm 0 5.5cm 0}]
{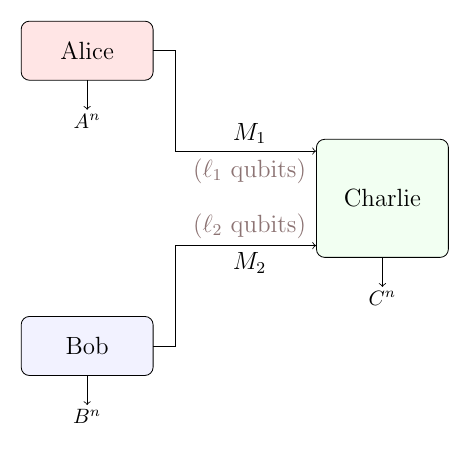} %
\caption{Multiple-access network.  Alice prepares the state of her output  $A^{n}$, as well as her ``quantum description",  $M_{1}$ consisting of $\ell_1$ qubits. 
Similarly, Bob prepares the state of his output  $B^{n}$, as well as   $M_{2}$ consisting of $\ell_2$ qubits.
They transmit $M_{1}$ and $M_2$ to Charlie. Upon receiving $M_{1}$ and $M_2$, Charlie encodes the output $C^{n}$. %
The objective is to simulate $n$ copies of a desired quantum state $\omega_{ABC}$, i.e., %
for the encoded state  $\widehat{\rho}_{A^{n}B^{n}C^n}$ to be arbitrarily close to  $\omega_{ABC}^{\otimes n}$.}
\label{Multiple_access_network}
\end{figure}

Consider the multiple-access network in Figure~\ref{Multiple_access_network}. Alice, Bob, and Charlie would like to simulate a pure state %
$\ket{\omega_{ABC}}^{\otimes n}$, where 
$\ket{\omega_{ABC}}\in\mathcal{H}_{A}\otimes\mathcal{H}_{B}\otimes\mathcal{H}_{C}$.
We simplify the notation
 and write $Q_1\equiv Q_{1,3}$ and
$Q_2\equiv Q_{2,3}$.
At first, Alice prepares the state of the quantum systems $A^{n}$ and $M_{1}$, and Bob prepares the states of the quantum systems $B^{n}$ and $M_{2}$. Alice and Bob send $M_{1}$ and $M_{2}$ to Charlie. Charlie then uses $M_1$ and $M_2$ to encode the system $C^{n}$. As in the previous settings, $M_{1}$ and $M_{2}$ are referred to as quantum descriptions, which are limited to the qubit transmission rates, $Q_1$ and $Q_2$, respectively.
\begin{definition}
\normalfont
A $(2^{\ell_{1}},2^{\ell_{2}},n)$ coordination code for the multiple-access network described in Figure~\ref{Multiple_access_network}, consists of two Hilbert spaces, $\mathcal{H}_{M_{1}}$ and $\mathcal{H}_{M_{2}}$, of dimensions
\begin{align} 
\mathrm{dim}(\mathcal{H}_{M_{j}})=2^{\ell_{j}}\;\text{for \ensuremath{j\in\{1,2\}} \,,} 
\end{align} 
and three encoding maps,
\begin{align}
\mathcal{E}_{A^{n}\to A^{n}M_{1}}	&:\Delta(\mathcal{H}_{A}^{\otimes n})\to\Delta(\mathcal{H}_{A}^{\otimes n}\otimes\mathcal{H}_{M_{1}})\,,
\\
\mathcal{F}_{B^{n}\to B^{n}M_{2}}	&:\Delta(\mathcal{H}_{B}^{\otimes n})\to\Delta(\mathcal{H}_{B}^{\otimes n}\otimes\mathcal{H}_{M_{2}})
\intertext{and}	
\mathcal{D}_{M_{1}M_{2}\to C^{n}}	&:\Delta(\mathcal{H}_{M_{1}}\otimes\mathcal{H}_{M_{2}})\to\Delta(\mathcal{H}_{C}^{\otimes n})\,,
\end{align}
\end{definition}
corresponding Alice, Bob, and Charlie, respectively.
The protocol has limited communication rates $Q_j$ for $j\in \{1,2\}$.
That is, Alice sends $\ell_1=nQ_1$ qubits to Charlie, while Bob sends $\ell_2=nQ_2$ qubits to Charlie. %

Alice and Bob apply the encoding maps,  preparing $\rho_{A^{n}M_{1} }^{(1)}\otimes \rho_{ B^n M_2}^{(1)}$, where
\begin{align}
\rho_{A^{n}M_{1}}^{(1)} &= \mathcal{E}_{A^{n}\to A^{n}M_{1}}
(\omega_{A}^{\otimes n})
\,,\;
\rho_{B^n M_2}^{(2)} = 
\mathcal{F}_{B^{n}\to B^{n}M_{2}}
(\omega_{B}^{\otimes n})
\,.
\label{Multiple_access_rho_1}
\end{align} 
As Charlie receives $M_{1}$ and $M_{2}$, he applies his encoding map, which yields the final state,
\begin{align}
\widehat{\rho}_{A^{n}B^{n}C^{n}} &= \left(\mathrm{id}_{A^{n}B^{n}}\otimes\mathcal{D}_{M_{1}M_{2}\to C^{n}}\right)(\rho_{A^{n}B^{n}M_{1}M_{2}})
\label{Multiple_access_rho_hat}
\end{align}
The ultimate goal of the coordination protocol is that the final state of $\widehat{\rho}_{A^{n}B^{n}C^{n}}$, is arbitrarily close to the desired state $\omega_{ABC}^{\otimes n}$.

\begin{remark}
\label{MAC_product_state}
Notice that since Charlie only acts on $M_{1}$ and $M_{2}$ which are encoded separately without coordination, we have $\widehat{\rho}_{A^{n}B^{n}}=\rho_{A^{n}}^{(1)}\otimes\rho_{B^{n}}^{(2)}$.
Therefore, it is only possible to simulate states $\omega_{ABC}$ such that $\omega_{AB}=\omega_A\otimes \omega_B$.
Since all purifications are isometrically equivalent
\cite[Theorem 5.1.1]{wilde2017quantum}
there exists an isometry $V_{C\to C_1 C_2}$ such that
\begin{align}
(\identity\otimes V_{C\to C_1 C_2})\ket{\omega_{A B C }}=
\ket{\phi_{AC_1}}\otimes \ket{\chi_{BC_2}}
\label{Equation:MAC_Isometry}
\end{align}
where $\ket{\phi_{AC_1}}$ and $\ket{\chi_{BC_2}}$ are purifications of $\omega_A$ and $\omega_B$, respectively.
If $\omega_{ABC}$ cannot be decomposed as in \eqref{Equation:MAC_Isometry}, then coordination is impossible in the multiple-access network.
\end{remark}

\begin{definition}
\normalfont
A rate pair $(Q_{1},Q_{2})$ is achievable, if for every $\varepsilon,\delta>0$ and a sufficiently large $n$, there exists a $(2^{n(Q_{1}+\delta)},2^{n(Q_{2}+\delta)},n)$ coordination code satisfying 
\begin{align}
\norm{\widehat{\rho}_{A^{n}B^{n}C^{n}}-\omega_{ABC}^{\otimes n}}_{1}	\leq\varepsilon \,.
\end{align} 
\end{definition}
\begin{remark}
The  resource inequality  for coordination in the multiple-access network is
\begin{align}
Q_1 [q\to q]_{A\to C}+Q_2 [q\to q]_{B\to C} \geq \left\langle \omega_{ABC} \right\rangle \,.
\end{align}
See resource definitions in Remark~\ref{Remark:Resource_Cascade}.
\end{remark}

The optimal coordination rates for the multiple-access network are provided below.
\begin{theorem}
\normalfont
\label{MAC_theorem}
Let $\ket{\omega_{ABC}}$ be a pure state as in \eqref{Equation:MAC_Isometry}.
Then, a rate pair $(Q_1,Q_2)$ for coordination in the  multiple-access network in Figure~\ref{Multiple_access_network} is achievable if and only if \begin{align}
Q_{1}       &\geq H(A)_\omega \,,\\
Q_{2}  &\geq H(B)_\omega \,.
\end{align}
\end{theorem}
The proof for Theorem~\ref{MAC_theorem} is provided in 
Section%
~\ref{Multiple_access_network_analysis}.

\begin{remark}
Consider the following trivial cases.
For $\ket{\omega_A}\otimes \ket{\omega_{BC}}$, we have $Q_1=0$, as coordination does not require any communication from Alice to Charlie. 
Similarly, for $\ket{\omega_{AC}}\otimes \ket{\omega_{B}}$, we have $Q_2=0$.
For example, if Alice and Charlie  simulate a maximally entangled qubit state $\omega_{AC}$, then the coordination region is $\mathcal{Q}_{\text{MAC}}(\omega)=\{(Q_1,0):Q_1\geq 1\}$.
Furthermore, if $\ket{\omega_{ABC}}=\ket{\omega_{AB}}\otimes \ket{\omega_{C}}$, then we also have $\omega_{AB}=\ket{\omega_A}\otimes \ket{\omega_B} $,
hence
coordination does not require  communication at all.
\end{remark}

\begin{remark}
\label{Remark: repeaters}
Consider a product of two maximally entangled qubit pairs, $\ket{\omega_{ABC_1 C_2}}=\ket{\phi_{AC_1}}\otimes \ket{\chi_{BC_2}}$, where $\mathcal{H}_C\equiv 
\mathcal{H}_{C_1}\otimes \mathcal{H}_{C_2}
$. The coordination capacity region is then $\mathcal{Q}_{\text{MAC}}(\omega)=\{(Q_1,Q_2):Q_i\geq 1\}$.
Now, suppose that Charlie performs a local Bell measurement on his qubits, $C_1$ and $C_2$, the entanglement is swapped such that $A$ and $B$ become maximally entangled.
We will discuss the implications for the application of quantum repeaters in Subsection~\ref{Subsection: Quantum repeaters}.
\end{remark}

\section{Nonlocal Games}
\label{Section:Nonlocal_games}
 In this section, we discuss the connection between quantum coordination and nonlocal games, focusing on the broadcast network. 
 We begin with a brief review on refereed games in Subsection~\ref{Subsection: Refereed games}. We explain how coordination is useful for a sequential game in Subsection~\ref{Subsection: Coordination in games}. We demonstrate the implications for 
 the CHSH game in Subsection~\ref{Subsection: CHSH}. 
\subsection{Nonlocal Correlations}
\label{Subsection: Refereed games}

\begin{figure}[t]
    \centering
    \hspace{0.5cm}
    \begin{minipage}[b]{0.3\linewidth} 
        \centering
        \includegraphics[scale=0.65,trim={5.3cm 0 5.5cm 0}]
        {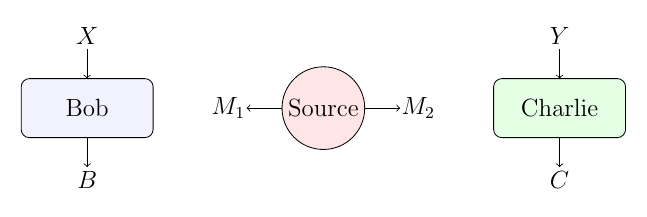} %
        \begin{center}
        \footnotesize{(a) Bell experiment setup.}
        \end{center}
    \end{minipage}
    \hspace{3.5cm}
    \begin{minipage}[b]{0.3\linewidth} 
        \centering
         \includegraphics[scale=0.65,trim={5.3cm -1cm 5.5cm 0}]
        {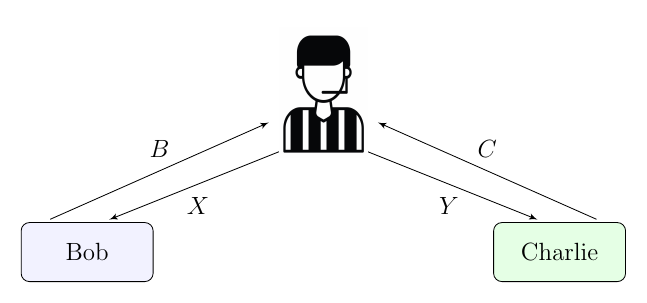} %
        \begin{center}
        \footnotesize{(b) Refereed game setting.}
        \end{center}
    \end{minipage}
    \caption{Bell experiments and refereed games. Figure~(a) describes a Bell experiment setup, consisting of a source, and two observers. The source distributes physical systems $M_1$ and $M_2$ to Bob and Charlie respectively. Bob and Charlie choose to perform measurements $X$ and $Y$ each on his system, yielding classical results $B$ and $C$ respectively. Figure~(b) describes a refereed game where a referee plays against both Bob and Charlie. The referee sends his questions $X$ and $Y$ to Bob and Charlie, they respond with answers $B$ and $C$ respectively. The round is won if the realization of the tuple $(X,Y,B,C)=(x,y,b,c)$ satisfies a specific condition set by the game rules.}
     \label{Figure: Bell experiments and refereed games}
\end{figure}

Nonlocal games are closely related to Bell experiments and quantum correlations \cite{brunner2014bell}.
In the typical setting for a Bell experiment, 
a source distributes two physical systems, $M_1$ and $M_2$, to two distant users. See Figure~\ref{Figure: Bell experiments and refereed games}~(a). 
Here, we refer to the users as Bob ($B$) and Charlie ($C$). Upon receiving $M_1$ and $M_2$, each chooses to perform a measurement from a certain set of measurements. Denote the measurements chosen by Bob and Charlie by $X$ and  $Y$, respectively. The measurements yield the respective outcomes, $B$ and $C$. Notice that $B$ and $C$ are classical in this setting. 

By the nature of quantum measurements \cite{wilde2017quantum}, the outcomes $B$ and $C$ may change from one run of the experiment to another, even when the same measurements $X$ and $Y$ are taken. The outcomes are governed by a conditional probability mass function $P_{BC|XY}(b,c|x,y)$, and can be estimated by running the experiment for a sufficient number of rounds. 
The function $P_{BC|XY}$ is also called a behavior, or, \emph{a correlation}.
In general, the correlation cannot necessarily be separated as $
P_{B|X}\times 
P_{C|Y}
$, even when the observers are remote. 
This does not necessarily imply a direct influence of one system on the other.

The notion of locality refers to a situation where past factors can be encapsulated in some random variable $U$, also referred to as a hidden variable \cite{einstein1935can,bell1966problem}, such that when taking it into account, the correlation between the outcomes is broken, i.e., 
\begin{align}
    P_{BC|XY}(b,c|x,y)=\int_{\text{supp}(U)}  {p_{U}(u)P_{B|XU}(b|x,u)P_{C|YU}(c|y,u) \,\text{d}u}\,.
    \label{Equation: locality}
\end{align}

The 
predictions of the quantum theory for certain settings
involving quantum entanglement do not follow the locality condition in \eqref{Equation: locality}.
Suppose that Bob and Charlie share a bipartite state $\rho_{M_1 M_2}$.
As they perform local measurements $\{F_{b|x}\,,\; b\in\mathcal{B}\}$ and 
$\{D_{c|y}\,,\; c\in\mathcal{C}\}$, they generate the following correlation: 
\begin{align}
    P_{BC|XY}(b,c|x,y)=\Tr\left[\left(F_{b|x}\otimes D_{c|y}\right)\rho_{M_{1}M_{2}}\right]\,.
\end{align}
The set of all quantum correlations is often denoted in the literature by $C_q$ \cite{slofstra2019set}.

One of the simplest experiments demonstrating nonlocal behavior is the CHSH  setting, named after Clauser, Horne, Shimony, and Holt \cite{clauser1969proposed}. Consider the Bell experiment setting shown in Figure~\ref{Figure: Bell experiments and refereed games}~(a), where the observers Bob and Charlie can only perform one of two measurements, $X,Y \in \{0,1\}$. The outcomes are limited to two values as well $B,C\in\{\pm 1\}$. Consider 
\begin{align}
    S=\langle B_0 C_0\rangle + \langle B_0 C_1\rangle +\langle B_1 C_0\rangle -\langle B_1 C_1\rangle\,.
    \label{Equation: S quantity}
\end{align}
where $\langle B_x C_y\rangle$ are the corresponding expectation values,
$
    \langle B_x C_y\rangle=\sum_{b,c\in \{\pm 1\}} bc \cdot P_{BC|XY}(b,c|x,y)
$, for $(x,y)\in\mathcal{X}\times\mathcal{Y}$. 

If the correlation $P_{BC|XY}$ satisfies the locality condition  in \eqref{Equation: locality}, then   $S\leq2$ must hold \cite{bell1964einstein}. However, in the quantum case,  this inequality may be violated. 
Suppose Bob and Charlie are  each provided with a qubit from an EPR pair $\ket{\Phi_{M_1M_2}}=\frac{1}{\sqrt{2}}\left(\ket{00}+\ket{11}\right)$. Denote the Pauli operators by $\left(\Sigma_1,\Sigma_2,\Sigma_3\right)$. Bob and Charlie choose their measurements depending on the values of $X$ and $Y$, respectively. 
If $X=0$, Bob measures the $\Sigma_3$ observable. Otherwise, if $X=1$, he measures the $\Sigma_1$ observable. As for Charlie, if $Y=0$, he measures the observable $\frac{-\Sigma_3-\Sigma_1}{\sqrt{2}}$, and if $Y=1$, he measures $\frac{\Sigma_3-\Sigma_1}{\sqrt{2}}$.
This 
yields $S=2\sqrt{2}>2$ (see \eqref{Equation: S quantity}), demonstrating the nonlocal nature of quantum entanglement. Based on 
this violation, quantum correlations cannot be explained using the  theory of classical hidden variables \cite{bell1964einstein}.   

\subsection{Refereed games}
Refereed games can be viewed as another representation of the Bell setting.
Specifically, 
consider the refereed game in Figure~\ref{Figure: Bell experiments and refereed games}~(b). 
The referee provides two  questions $X\in \mathcal{X}$ and $Y\in \mathcal{Y}$, according to some probability distribution $p_{X,Y}$.
He sends $X$ to the first player (Bob), and $Y$ to the second (Charlie).
Upon receiving their question,  Bob and Charlie respond with classical answers $B\in \mathcal{B}$ and $C\in \mathcal{C}$, respectively. We note that the alphabets $\mathcal{X}$, $\mathcal{Y}$, $\mathcal{B}$, and $\mathcal{C}$ are assumed to be finite. 
The referee 
decides that the game is won if the realization of the tuple $(X,Y,B,C)$ satisfies a specific condition $\mathscr{W}$, set by the rules of the game. This condition is represented by an indicator function,
\begin{align}
    V(x,y,b,c) =
    \begin{cases} 
      1 &  \text{If $(x,y,b,c)$ satisfy $\mathscr{W}$}\,, \\
      0 & \text{otherwise }\,.
\end{cases}
    \label{Equation: Indicator}
\end{align}

We refer to the procedure above 
as a single-shot game. 
 We now discuss the game implementation and rules.

$\,$

\subsubsection{Resources}
As in the Bell setting,  a source distributes correlated physical systems before the procedure begins (see Figure~\ref{Figure: Bell experiments and refereed games}~(a)).
Here, we refer to the source of the correlation resources as \emph{Alice}. 

$\,$

\subsubsection{Strategy}
Before the game starts, i.e., before the referee has chosen his questions, Alice, Bob, and Charlie  meet and agree on a game strategy and the required correlation resources. 
The optimal game strategy and the required correlations for the strategy implementation depend on the game rules. 

$\,$

\subsubsection{No signaling}
During  the course of the game, Bob and Charlie cannot communicate with each other. 
They can, however, exploit the correlation resources in order to coordinate their answers  through quantum measurements.

$\,$

We can also give an equivalent description of the game implementation in terms of three phases.
In Phase~1, the source (Alice) distributes the correlation resources, $M_1$ and $M_2$,  between the players (Bob and Charlie). 
In Phase~2, the referee generates the question pair $(x,y)$ according to $p_{XY}$, and sends $x$ and $y$ them to Bob and Charlie, respectively. 
In Phase~3, upon receiving their questions, Bob and Charlie produce their answers, $B$ and $C$. Once the referee is informed, he decides whether the game is won. 
We refer to this description as a single shot game. 

The winning probability is thus
\begin{align}
\pi(P_{BC|XY})=\sum_{(x,y,b,c)\in\mathcal{X}\times\mathcal{Y}\times\mathcal{B}\times\mathcal{C}} p_{XY}(x,y)P_{BC|XY}(b,c|x,y) \cdot V(x,y,b,c)\,.
\label{Equation: Winning probability}
\end{align}
The performance depends directly on the correlation $P_{BC|XY}$ that Alice, Bob, and Charlie simulate as a consequence of the three phases above. 
 For example, in the CHSH game, 
the winning condition is
$
    x \wedge y = b \oplus c \,.
$, where $x,y,b,c\in \{0,1\}$. 
$\pi(P_{BC|XY})=\frac{1}{2}\left(1+\frac{S}{4}\right)$.
 Classical strategies may generate correlations $P_{BC|XY}$ such that  $S\leq 2$ (see \eqref{Equation: S quantity}), hence, the game can be won with probability  $\pi(P_{BC|XY})\leq 0.75$. Whereas, entanglement allows for $S=2\sqrt{2}$, for which $\pi(P_{BC|XY})= 0.8535$. 

\subsection{Sequential game}
\begin{figure}[t]
    \centering
    \begin{minipage}[b]{0.3\linewidth} 
        \centering
        \includegraphics[scale=0.65,trim={5.3cm 0cm 5.5cm 0}]{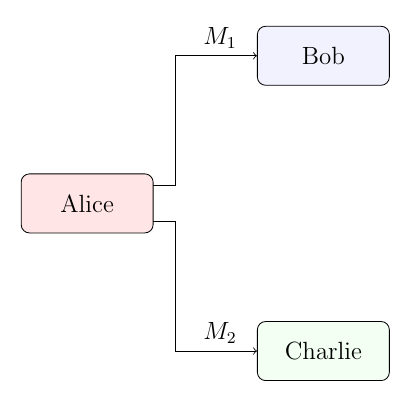} 
        \vspace{0.25cm}
        \begin{center}
        \footnotesize{(a) Phase~1.} 
        \end{center}
    \end{minipage}
    \hspace{0.5cm}
    \begin{minipage}[b]{0.3\linewidth} 
        \centering
        \includegraphics[scale=0.65,trim={5.3cm 0 5.5cm 0}]{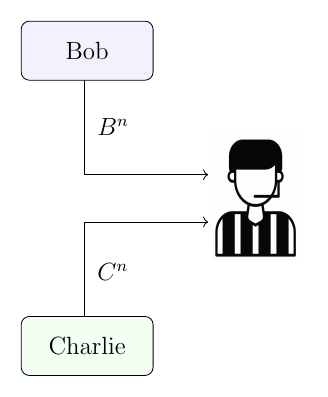} 
        \vspace{0.25cm}
        \begin{center}
        \footnotesize{(b) Phase~2.}
        \end{center}
    \end{minipage}
    \hspace{0.5cm}
    \begin{minipage}[b]{0.3\linewidth} 
        \centering
        \includegraphics[scale=0.65,trim={5.3cm 0 5.5cm 0}]{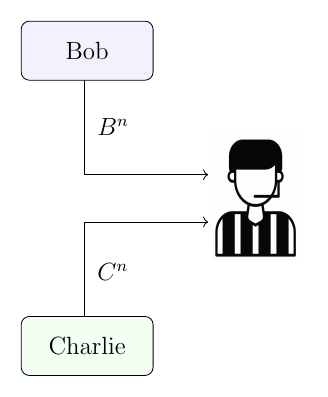} 
        \vspace{0.25cm}
        \begin{center}
        \footnotesize{(c) Phase~3.}
        \end{center}
    \end{minipage}
    \caption{Implementation of a refereed game in three phases:  
(a) 
In Phase~1, the source (Alice) distributes the correlation resources  between the players (Bob and Charlie). 
(b) 
In Phase~2, the referee generates the questions, and sends  the respective question to Bob and Charlie.
(c)
In Phase~3, upon receiving their questions, Bob and Charlie produce their answers and inform the referee. Once the referee is informed, he decides whether the game is won.  
}
    \label{Figure: Nonlocal games}
\end{figure}
We now introduce a sequential version of the refereed game, see Figure~\ref{Figure: Nonlocal games}. 
In Phase~1, the source (Alice) distributes the correlation resources, $M_1$ and $M_2$,  between the players (Bob and Charlie).  
In   Phase~2, the referee generates a sequence of $n$ independent question pairs $(x_i,y_i)$ according to $p_{XY}$,  and sends $x^n$ and $y^n$ to Bob and Charlie, respectively. 
In Phase~3, the players produce their responses. Bob and Charlie choose their measurements depending on  $x^n$ and $y^n$, respectively.
Then, they perform  their respective measurements on $M_1$ and $M_2$. They   send the measurement outcomes $b^n$ and $c^n$, respectively,  to the referee. The worst-case winning probability is thus
\begin{align}
\pi_n(\widehat{P}_{B^nC^n|X^nY^n})=\min_{i\in [n]} \sum_{(x^n,y^n,b^n,c^n)\in\mathcal{X}^n\times\mathcal{Y}^n\times\mathcal{B}^n\times\mathcal{C}^n} p^n_{XY}(x^n,y^n)\widehat{P}_{B^nC^n|X^nY^n}(b^n,c^n|x^n,y^n) \cdot V(x_i,y_i,b_i,c_i)\,,
\end{align}
Notice that if we simulate a product correlation, i.e.,
$\widehat{P}_{B^nC^n|X^nY^n}\approx P_{BC|XY}^n$, then
\begin{align}
\pi_n(\widehat{P}_{B^nC^n|X^nY^n})\approx \pi(P_{BC|XY}) \,.
\end{align}

\subsection{Coordination as part of a game strategy}
\label{Subsection: Coordination in games}


We now present the connection between  quantum coordination and refereed games explicitly. We now insert a broadcast coordination scheme into the game strategy. We consider the special case where $B$ and $C$ are classical, while
$A$ is null (say, 
$\mathrm{dim}(\mathcal{H}_A)=1$).

Consider the sequential game setup described in Figure~\ref{Figure: Nonlocal games}. 
In Phase 1, the source (Alice) prepares the quantum resources $M_1$ and $M_2$ using the coordination encoding map 
$\mathcal{E}$. She then distributes the resources 
between the respective players (Bob and Charlie), using noiseless quantum links at rates $Q_1$ and $Q_2$. 
In Phase 2, the referee chooses question sequences $X^n$ and $Y^n$ that have no correlation with the quantum resources, as in 
the broadcast network model. 
In Phase 3, Bob and Charlie use the encoding measurements $\mathcal{F}_{X^n M_1\to B^n}$ and
$\mathcal{D}_{Y^n M_2\to C^n}$. 
They obtain $B^n$ and $C^n$ as measurement outcomes and inform the referee. 

This coordination strtegy generates a classical-correlation state,
$\widehat{\rho}_{X^n Y^n B^n C^n}\approx \omega_{XYBC}^{\otimes n}$, where
\begin{align}
\omega_{XYBC}=\sum_{(x,y,b,c)\in\mathcal{X}\times\mathcal{Y}\times\mathcal{B}\times\mathcal{C}} p_{XY}(x,y)P_{BC|XY}(b,c|x,y)
\ketbra{x,y,b,c} \,,
\end{align}
which leads to a minimal winning probability $\pi(P_{BC|XY})$ (see \eqref{Equation: Winning probability}).


Let $\mathscr{S}(\gamma)$ denote the set of correlations $P_{BC|XY}$ that win the game with probability $\gamma$.
Based on our results,  the game can be won with minimal probability $\gamma$ if and only if
Alice can send qubits to Bob and Charlie at rates $Q_1$ and $Q_2$ that satisfy the constraints in Theorem~\ref{Theorem_Broadcast}
with respect to some correlation $P_{BC|XY}\in \mathscr{S}(\gamma)$.
Being able to generate entanglement between Bob and Charlie, can provide an advantage by inducing quantum correlations stronger than their classical counterparts, hence allowing for higher winning probabilities (see Subsection~\ref{Subsection: Refereed games}).

\subsection{Example: The CHSH game}
\label{Subsection: CHSH}
%
A well known game demonstrating the advantage of nonlocal correlations is the CHSH game. Suppose that the players first simulate the following state using a broadcast coordination code: 
%
\begin{align}
    \ket{\omega^{(x,y)}
    }&=\sqrt{\alpha_{x,y}}\ket{00}+\sqrt{1-\alpha_{x,y}}\ket{11}\,,
\end{align}
where $\alpha_{x,y}$ are given parameters in $[0,1]$, for 
$(x,y)\in \mathcal{X}\times\mathcal{Y}$. Applying the same measurement strategy as in the CHSH experiment in Subsection~\ref{Subsection: Refereed games}, we obtain a correlation $P$  such that 
the winning probability is given by
\begin{align}
    \pi^{\text{CHSH}}(P)=\frac{1}{16}\sum_{x,y\in \{0,1\}}\left[\frac{1+2\sqrt{2}}{\sqrt{2}}+\sqrt{2\alpha_{x,y}(1-\alpha_{x,y})}\right]\,.
\end{align}
\begin{figure}[t]
    \centering
    \includegraphics[scale=0.22,trim={5.3cm 0 5.5cm 0}]{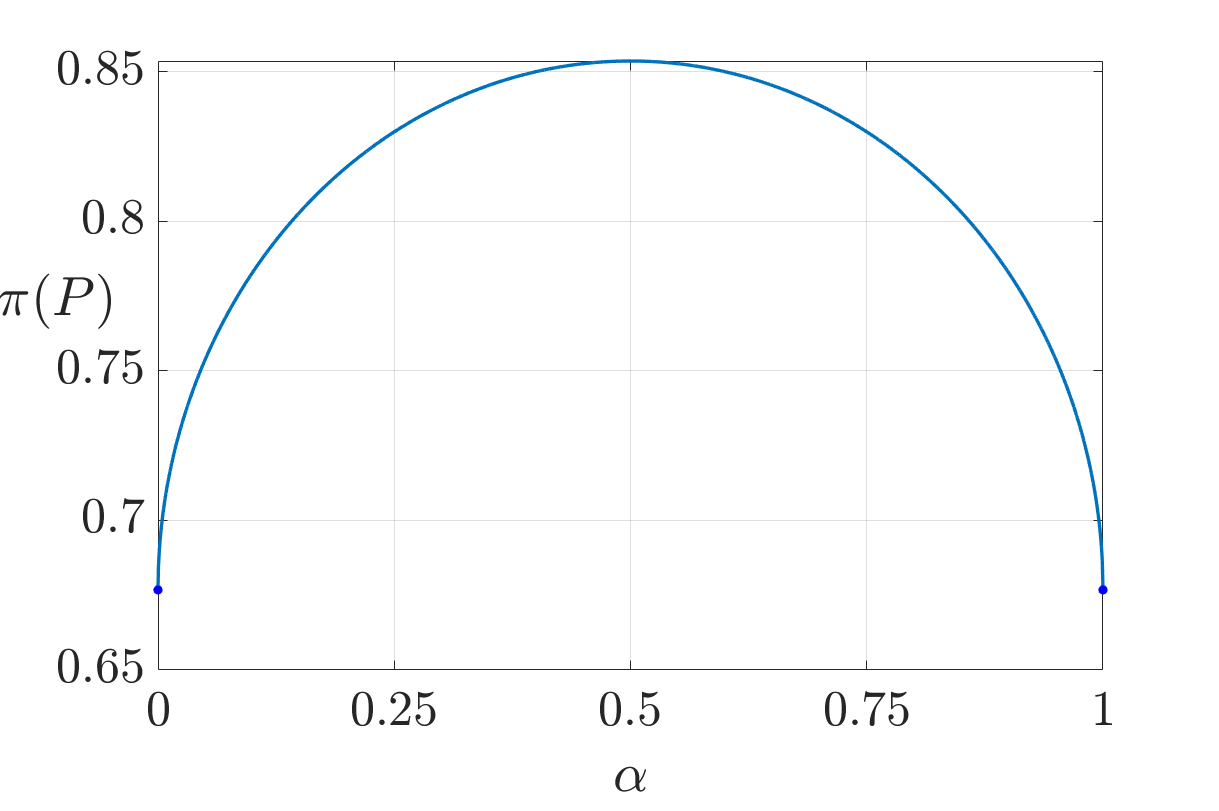}
    \centering
    \caption{ Winning probability as a function of $\alpha$.}  
    \label{Figure: Winning probability}
\end{figure}
For
$\alpha\equiv \frac{1}{2}$, i.e., with maximal Bell violation \cite{cirel1980quantum}, we get a maximal winning probability of $\pi^{\text{CHSH}}(P)
= 0.8535$.
For 
$\alpha\equiv 0$, when there is no correlation, we have $\pi^{\text{CHSH}}(P)=0.6767$. In this case, the CHSH measurement strategy is even worse that the best classical strategy, for which $\pi^{\text{CHSH}}(P)=0.75$. 

By Theorem~\ref{Theorem_Broadcast}, Phase 1 requires the communication rates $Q_1\geq 
\frac{1}{2}H_2\left( 
 \frac{1}{2}\left( \alpha_{0,0}+\alpha_{0,1} \right)\right)+\frac{1}{2}H_2\left( 
 \frac{1}{2}\left( \alpha_{1,0}+\alpha_{1,1} \right)\right)$ and $Q_2\geq 
 \frac{1}{2}H_2\left( 
 \frac{1}{2}\left( \alpha_{0,0}+\alpha_{1,0} \right)\right)+\frac{1}{2}H_2\left( 
 \frac{1}{2}\left( \alpha_{0,1}+\alpha_{1,1} \right)\right)$, where $H_2(\cdot)$ is the binary entropy function. In particular, for a constant parameter, $\alpha_{x,y}=\alpha$ for all $ x,y$, we have $Q_j\geq H_2(\alpha)$. 
There is a threshold value $\alpha^*$ for which the CHSH measurement strategy has the same performance as the best classical strategy. 
Specifically, we obtain a Bell violation provided that $\alpha_{x,y}>0.04491$ for all $(x,y)\in\mathcal{X}\times\mathcal{Y}$.
The winning probability for a constant parameter $\alpha$, is shown in Figure~\ref{Figure: Winning probability}. Since the gradient is unbounded near 
$\alpha=0$, 
even a small amount of entanglement can have a significant effect on the winning probability. As we approach 
$\alpha=\frac{1}{2}$, the gradient diminishes. 
 The Bell violation threshold requires
$Q_j\geq H_2(\alpha^*)= 0.2643$.

To summarize, we have discussed 
the notion of Bell experiments and their direct connection to 
the simulation of nonlocal correlations. 
We then discussed refereed games in the standard single-shot form and the sequential form. We have shown that coordination in the broadcast network can be viewed as the overall game strategy, i.e., the preparation of the pre-shared resources (Phase 1) and the measurement (Phase 3). In this sense, coordination is the enabler of quantum strategies that achieve higher  
winning probabilities compared to classical ones.

Quantum coordination can be useful for generating nontrivial correlations in other types of games as well. For example, in pseudo-telepathy games, quantum strategies guarantee winning with probability $1$. 
One example is the magic square game \cite{brassard2005quantum}, where $(X,Y)$ are the coordinates of a cell in the square,
and the players win the game if they can provide $3$ bits each that satisfy a parity condition. 
In this case, the game can be won with $Q_j=2$ qubits per question, for each player.
Slofstra and Vidick \cite{slofstra2018entanglement} presented a game where coordination of a correlation that could win with probability $(1-e^{-T})$ requires
$Q_j\propto T$ qubits per question.

\section{Cascade Network Analysis} %
\label{Cascade_network_analysis}
We prove the rate characterization in Theorem~\ref{Theorem_Cascade}. 
Consider the cascade network in Figure~\ref{Cascade_network}.  
\subsection{Achievability proof}
The proof for the direct part exploits the state redistribution result by Yard and Devetak in \cite{Yard_Devetak_2009}. We first describe the state redistribution problem.
Consider two parties, Alice and Bob. Let their systems be described by the joint state  $\psi_{A B G}$, where  $A$ and $B$ belong to Alice, and  $G$ belongs to Bob. Let the state $\ket{\psi_{A B G R}}$ be a purification of $\psi_{A B G}$. Alice and Bob would like to redistribute the state $\psi_{A B G}$ such that  $B$ is transferred from Alice to Bob. Alice can send quantum description systems at rate $Q$ and they share maximally entangled pairs of qubits at a rate $E$.
\begin{theorem}[State Redistribution \cite{Yard_Devetak_2009}]
\normalfont
\label{Yard_and_Devetak}
The optimal rates for state redistribution of $\ket{\psi_{A G B R}}$ with rate-limited entanglement  are
\begin{align}
    Q	&\geq\frac{1}{2}I(B;R|G)_{\psi} \,,
    \\
    Q+E	&\geq H(B|G)_{\psi} \,.
\end{align}
\end{theorem}
We go back to the coordination setting for the cascade network (see Figure~\ref{Cascade_network}).
Alice, Bob, and Charlie would like to simulate the joint state $\omega_{ABC}^{\otimes n}$. Let $\ket{\omega_{ABCR}}$ be a purification.
Suppose that Alice prepares the desired state $\ket{\omega_{A\bar{B}\bar{C}\bar{R}}}^{\otimes n}$ locally in her lab, where $\bar{B}^{n}$, $ \bar{C}^{n}$, and $ \bar{R}^{n}$ are her ancillas. Let $\varepsilon>0$ be arbitrarily small. By the state redistribution theorem, Theorem~\ref{Yard_and_Devetak}, Alice can transmit $\bar{B}^{n}\bar{C}^{n}$  to Bob at communication rate $Q_1$ and entanglement rate $E_1$, provided that 
\begin{align}
Q_{1}	&\geq\frac{1}{2}I(\bar{B}\bar{C};\bar{R})_{\omega}=\frac{1}{2}I(BC;R)_{\omega} \,,
\\
Q_{1}+E_{1}	&\geq H(\bar{B}\bar{C})_{\omega}=H(BC)_{\omega}
\end{align}
(see \cite{Yard_Devetak_2009}).
That is, there exist a  bipartite state $\Psi_{T_A T_B'}$ and
encoding maps, $\mathcal{E}_{\bar{B}^{n}\bar{C}^{n}T_A\to M_{1}}^{(1)}$ and  $\mathcal{F}_{M_{1}T_{B}'\to B^{n}\widetilde{C}^{n}}^{(1)}$, such that 
\begin{align}\norm{\tau_{\bar{R}^{n}A^{n}B^{n}\widetilde{C}^{n}}^{(1)}-\omega_{RABC}^{\otimes n}}_{1}\leq\varepsilon,
\label{Yard_1_cascade}
\end{align}
for sufficiently large $n$, where 
\begin{align} 
\tau_{\bar{R}^{n}A^{n}B^{n}\widetilde{C}^{n}}^{(1)} = \left[\mathrm{id}_{\bar{R}^{n}A^{n}} \otimes \mathcal{F}_{M_{1}T_{B}'\to B^{n}\widetilde{C}^{n}}^{(1)} \circ \left(\mathcal{E}_{\bar{B}^{n}\bar{C}^{n}T_{A}\to M_{1}}^{(1)} \otimes \mathrm{id}_{T_{B}'}\right)\right] 
\left(\omega_{RABC}^{\otimes n} \otimes \Psi_{T_{A}T_{B}'}\right)\,. \label{Tau_1_cascade} 
\end{align}
Similarly,  $\bar{C}^{n}$  can be compressed and transmitted with rates
\begin{align}
Q_{2}	&\geq\frac{1}{2}I(\bar{C};A\bar{R})_{\omega}=\frac{1}{2}I(C;AR)_{\omega} \,,
\\
Q_{2}+E_{2}	&\geq H(\bar{C})_{\omega}=H(C)_{\omega} \,,
\end{align}
by Theorem~\ref{Yard_and_Devetak}.
Namely, there exists a bipartite state $\Theta_{T_B'' T_C}$ and  encoding maps, $\mathcal{F}_{\bar{C}^{n} T_B''\to M_{2}}^{(2)}$ and $\mathcal{D}_{ M_{2} T_C\to C^{n}}^{(2)}$, such that 
\begin{align}
\norm{\tau_{\bar{R}^{n}A^{n}B^{n}C^{n}}^{(2)}-\omega_{\bar{R}ABC}^{\otimes n}}_{1}\leq\varepsilon,
\label{Yard_2_cascade} 
\end{align} 
where
\begin{align} 
\tau_{\bar{R}^{n}A^{n}B^{n}C^{n}}^{(2)}=%
 \left[
 \left(\mathrm{id}_{\bar{R}^{n} A^{n} \bar{B}^n }\otimes 
 \mathcal{D}_{M_{2}T_C\to C^{n}}^{(2)}
\right)\circ
 \left(
\mathcal{F}_{\bar{C}^{n}T_B''\to M_{2}}^{(2)}\otimes \mathrm{id}_{T_C} 
 \right)
 \right]
 \left(\omega_{\bar{R} A\bar{B}\bar{C}}^{\otimes n}\otimes\Theta_{T_B'' T_C}\right) \,.
\label{Tau_2_cascade} 
\end{align}

The coding operations for the cascade network are described below.
\paragraph*{Encoding}
\begin{enumerate}[A)]
\item%
Alice prepares $\ket{\omega_{A\bar{B}\bar{C}\bar{R}}}^{\otimes n}$ locally. She applies $\mathrm{id}_{\bar{R}^{n}A^{n}}\otimes\mathcal{E}_{\bar{B}^{n}\bar{C}^{n}T_A\to M_{1}}^{(1)}$, and sends $M_{1}$ to Bob.

\item%
As Bob receives $M_{1}$, he applies
\begin{align}
\mathcal{F}_{M_{1}T_{B}'T_{B}''\to B^{n}M_{2}}  \equiv \left(\mathrm{id}_{B^{n}}\otimes\mathcal{F}_{\widetilde{C}^{n}T_{B}''\to M_{2}}^{(2)}\right) \circ \mathcal{F}_{M_{1}T_{B}'\to B^{n}\widetilde{C}^{n}}^{(1)}\,. 
\label{Bob_cascade}
\end{align}

\item%
Charlie receives $M_{2}$ from Bob and applies $\mathcal{D}_{M_{2}T_{C}\to C^{n}}^{(2)}$.
\end{enumerate}

\paragraph*{Error analysis} 
We trace out the reference system $R$ and write the analysis with respect to the reduced states. The joint state after Alice's encoding is
\begin{align}\rho_{A^{n}M_{1}T_{B}'}^{(1)}=\left[\mathrm{id}_{A^{n}}\otimes\mathcal{E}_{\bar{B}^{n}\bar{C}^{n}T_{A}\to M_{1}}^{(1)}\otimes\mathrm{id}_{T_{B}'}\right](\omega_{A\bar{B}\bar{C}}^{\otimes n}\otimes\Psi_{T_{A}T_{B}'})\,. 
\end{align}
After Bob applies his encoder, this results in 
\begin{align}
\rho_{A^{n}B^{n}M_{2}T_{C}}^{(2)} &= \left[\left(\mathrm{id}_{A^{n}B^{n}} \otimes \mathcal{F}_{\widetilde{C}^{n}T_{B}''\to M_{2}}^{(2)} \otimes \mathrm{id}_{T_{C}}\right) \circ  \left(\mathrm{id}_{A^{n}} \otimes \mathcal{F}_{M_{1}T_{B}'\to B^{n}\widetilde{C}^{n}}^{(1)} \otimes \mathrm{id}_{T_{B}''\,T_{C}}\right) \right](\rho_{A^{n}M_{1}T_{B}'}^{(1)} \otimes \Theta_{T_{B}''\,T_{C}}) \nonumber \\
& = \left(\mathrm{id}_{A^{n}B^{n}} \otimes \mathcal{F}_{\widetilde{C}^{n}T_{B}''\to M_{2}}^{(2)} \otimes \mathrm{id}_{T_{C}}\right) (\tau_{A^{n}B^{n}\widetilde{C}^{n}}^{(1)} \otimes \Theta_{T_{B}''\,T_{C}})
\end{align}
by \eqref{Bob_cascade}, and
based on the definition of $\tau^{(1)}$ in
\eqref{Tau_1_cascade}. According to \eqref{Yard_1_cascade}, $\tau^{(1)}$ and $\omega^{\otimes n}$ are close in trace distance. By trace monotonicity under quantum channels, we have
\begin{align}
\norm{\rho_{A^{n}B^{n}M_{2}T_{C}}^{(2)}-\left(\mathrm{id}_{A^{n}B^{n}}\otimes\mathcal{F}_{\widetilde{C}^{n}T_{B}''\to M_{2}}^{(2)}\otimes \mathrm{id}_{T_{C}}\right)(\omega_{AB\widetilde{C}}^{\otimes n}\otimes \Theta_{T_B''\,T_C})}_{1}\leq\varepsilon.
\label{Yard_1_Consequence_A}
\end{align}
As Charlie receives $M_{2}$ and encodes, the final state, at the output of the cascade network, is given by
\begin{align} 
\widehat{\rho}_{A^{n}B^{n}C^{n}}=\left[\mathrm{id}_{A^{n}B^{n}}\otimes\mathcal{D}_{M_{2}T_{C}\to C^{n}}^{(2)}\right](\rho_{A^{n}B^{n}M_{2}T_{C}}^{(2)})\,.
\end{align}
Once more, by trace monotonicity, 
\begin{align}
\norm{\widehat{\rho}_{A^{n}B^{n}C^{n}}-\tau_{A^{n}B^{n}C^{n}}^{(2)}}_{1}\leq\varepsilon\,.
\label{Yard_1_Consequence_B}
\end{align}
(see \eqref{Yard_2_cascade} and \eqref{Tau_2_cascade}).
Thus, using \eqref{Yard_2_cascade}, \eqref{Yard_1_Consequence_B}, and the triangle inequality, we have
\begin{align} 
\norm{\widehat{\rho}_{A^{n}B^{n}C^{n}}-\omega_{ABC}^{\otimes n}}_{1}	&\leq\norm{\tau_{A^{n}B^{n}C^{n}}^{(2)}-\omega_{ABC}^{\otimes n}}_{1}+\norm{\widehat{\rho}_{A^{n}B^{n}C^{n}}-\tau_{A^{n}B^{n}C^{n}}^{(2)}}_{1}
\nonumber
\\
&\leq2\varepsilon\,.
\end{align}
This completes the achievability proof for the cascade network.
\subsection{Converse proof}
We now prove the converse part for Theorem~\ref{Theorem_Cascade}.
Recall that in the cascade network, each party shares entanglement with their nearest neighbor a priori, i.e.,
Alice and Bob share $\ket{\Psi_{T_A T_B'}}$, while Bob and Charlie share $\ket{\Theta_{T_B'' T_C}}$ %
(see Figure~\ref{Cascade_network} in subsection \ref{Cascade_subsection}).
Alice applies an encoding map $\mathcal{E}_{\bar{A}^n T_A \to A^{n}M_{1}}$ on her part, and sends the output $M_1$ to Bob.  %
As Bob receives $M_{1}$, %
he encodes %
using a map $\mathcal{F}_{M_{1}T_B' T_B'' \to B^{n}M_{2}}$,
and sends $M_{2}$. As Charlie receives $M_{2}$, he applies an encoding channel $\mathcal{D}_{M_{2}T_C\to C^{n}}$.
Suppose that Alice prepares the state 
$\ket{\omega_{R\bar{A}\bar{B}\bar{C}}}^{\otimes n}$
locally, and then encodes as explained above.
The protocol can be described through the following relations:
\begin{align} 
\rho_{R^n A^{n} M_{1}T_B'}^{(1)}&=
(\mathrm{id}_{R^n}\otimes \mathcal{E}_{ \bar{A}^n T_A \to A^{n}M_{1}}\otimes \mathrm{id}_{T_B'})(\omega_{R\bar{A}}^{\otimes n}\otimes \Psi_{T_A T_B'} ) \,, 
\label{Cascade_converse_rho_1}
\\
\rho_{R^n A^{n}B^{n}M_{2}T_C}^{(2)}&=(\mathrm{id}_{\bar{R}^n A^{n}} \otimes\mathcal{F}_{M_{1}T_B' T_B''\to B^{n}M_{2}}\otimes 
\mathrm{id}_{T_C})(\rho_{R^n A^{n}M_{1} T_B'}^{(1)}
\otimes
\Theta_{T_B''\, T_C})\,,
\label{Cascade_converse_rho_2}
\\
\widehat{\rho}_{R^n A^{n}B^{n}C^{n}}&=(\mathrm{id}_{R^n A^{n}B^{n}}\otimes\mathcal{D}_{M_{2}T_C\to C^{n}})\left(\rho_{R^n A^{n}B^{n}M_{2}T_C}^{(2)}\right)\,.
\label{Cascade_converse_rho_hat}
\end{align}

Let $\left(Q_{1},Q_{2},E_{1},E_{2}\right)$ be achievable rate tuple for coordination in the cascade network. Then, there exists a sequence of  codes %
such that 
\begin{align}
    \left\Vert \widehat{\rho}_{R^n A^{n}B^{n}C^{n}} - \omega_{RABC}^{\otimes n}\right\Vert _{1}\leq\varepsilon_n 
    \label{Cascade_converse_error}
\end{align} 
where $\varepsilon_n \to 0$ as $n\to \infty$.
\begin{figure}[bt]
\center
\includegraphics[scale=0.8,trim={5.3cm 0 5.5cm 0}]
{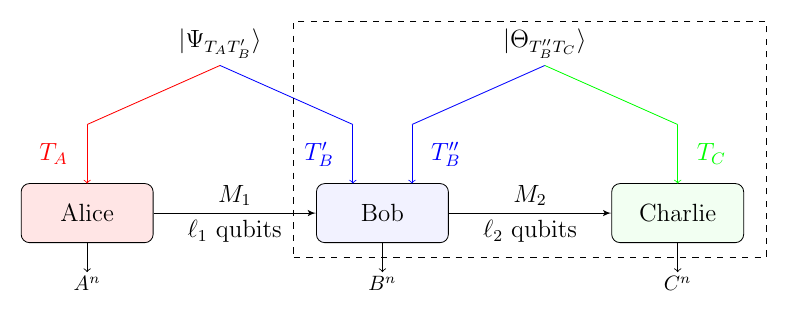} %
\caption{At first, we treat the encoding operation of Bob and Charlie as a black box.
}
\label{Cascade_network_with_a_box}
\end{figure}
Consider Alice's communication and entanglement rates, $Q_1$ and $E_1$. At this point, we may view the entire encoding operation of Bob and Charlie as a black box whose input and output are $(M_1,T_B')$ and $(B^n,C^n)$, respectively, as illustrated in Figure~\ref{Cascade_network_with_a_box}.
Now,
\begin{align}
2n(Q_{1}+E_{1})	&=2\left[\log\dim(\mathcal{H}_{M_{1}})+\log\dim(\mathcal{H}_{T_{B}'})\right]
\nonumber\\
& {\geq}I(M_{1}T_{B}';A^{n}R^{n})_{\rho^{(1)}}
\label{Equation:Converse:Q1E1_rho1}
\end{align}
since the quantum mutual information satisfies  $I(A;B)_\rho\leq 2\log \mathrm{dim}(
\mathcal{H}_A
)$ in general.
Therefore, by the data processing inequality,
\begin{align}
I(M_{1}T_{B}';A^{n}R^{n})_{\rho^{(1)}}
\,&{\geq}I(B^n C^{n};A^{n}R^{n})_{\widehat{\rho}}
\nonumber\\
\,& {\geq}I(B^n C^{n};A^{n}R^{n})_{\omega^{\otimes n}}-n\alpha_{n}
\nonumber\\
\,&=
n[I(BC;A R )_{\omega}-\alpha_{n}] \,,
\label{Equation:Converse:Q1E1_DPI}
\end{align}
where $\alpha_{n}\to0$ when $n\to\infty$.
The second inequality follows from \eqref{Cascade_converse_error} and the Alicki-Fannes-Winter (AFW) inequality \cite{AFW_Winter_2016} (entropy continuity).
Since $\ket{\omega_{RABC}}$ is pure, we have 
$I(BC;A R )_{\omega}=2H(BC)_{\omega}$ \cite[Th. 11.2.1]{wilde2017quantum}.
Therefore, combining \eqref{Equation:Converse:Q1E1_rho1}-\eqref{Equation:Converse:Q1E1_DPI}, we have 
\begin{align}
Q_1+E_1\geq H(BC)_{\omega}-\frac{1}{2}\alpha_{n} \,.
\end{align}

To show the bound on $Q_1$, observe that a lower bound on the communication rate with unlimited entanglement resources also holds with limited resources.
 Therefore, the bound 
$%
 Q_{1}\geq \frac{1}{2} I(BC;R)_{\omega}
$ %
follows from the entanglement-assisted capacity theorem due to Bennett et al. \cite{bennett2002entanglement}.
It is easier to see this through  resource inequalities, following the arguments in \cite{Qreverse_Shannon_2009}.
If the entanglement resources are unlimited, then the coordination code achieves
\begin{align}
Q_1\, [q\to q]_{A\to BC} &\geq \left\langle \omega_{RBC} \right\rangle
\nonumber\\
&\equiv \left\langle \trace_A:\omega_{RABC} \right\rangle
\nonumber\\
&\geq \frac{1}{2} I(BC;R)_{\omega}\, [q\to q]_{A\to BC}
\end{align}
where the resource units $[q\to q]$, $[q q]$, and $\left\langle \omega_{ABC} \right\rangle$ represent a single use of a noiseless qubit channel, an EPR pair, and the desired state $\omega_{ABC}$, respectively, while the unit resource $\left\langle \mathcal{N}_{A\to B}: \rho \right\rangle$ indicates a simulation of the channel output from $\mathcal{N}_{A\to B}$ with respect to the input state $\rho$.
The last inequality holds by \cite{bennett2002entanglement,Qreverse_Shannon_2009}.

Similarly, we bound Bob's communication and entanglement rates as follows,
\begin{align}
2n(Q_{2}+E_{2})	&=2\left[\log\dim(\mathcal{H}_{M_{2}})+\log\dim(\mathcal{H}_{T_{B}''})\right]
\\
&{\geq}I(M_{2}T_{C};A^{n}B^{n}R^{n})_{\rho^{(2)}}
\\
&{\geq}I(C^{n};A^{n}B^{n}R^{n}T_{B}'')_{\widehat{\rho}}
\\
&{\geq} n[I(C;A B R )_{\omega}-\beta_n]
\\
&=n[2H(C)_{\omega}-\beta_n]
\end{align}
where $\beta_{n}\to0$ when $n\to\infty$. As before, the last inequality follows from
  \eqref{Cascade_converse_error} and the AFW inequality \cite{AFW_Winter_2016}. Hence, 
\begin{align}
Q_2+E_2\geq H(C)_{\omega}-\frac{1}{2}\beta_{n} \,.
\end{align}
Furthermore,
\begin{align}
Q_2\, [q\to q]_{B\to C} &\geq \left\langle \omega_{RAC} \right\rangle
\nonumber\\
&\equiv \left\langle \trace_{B}:\omega_{RABC} \right\rangle
\nonumber\\
&\geq \frac{1}{2} I(C;AR)_{\omega}\, [q\to q]_{ B\to C}
\end{align}
which implies $Q_2\geq \frac{1}{2} I(C;AR)_{\omega}$.

This completes the proof of Theorem~\ref{Theorem_Cascade} for the cascade network.

\section{Broadcast Analysis%
}
\label{Broadcast_network_analysis}
We prove the rate characterization in Theorem~\ref{Theorem_Broadcast}. 
Consider the broadcast network in Figure~\ref{Broadcast_network}. 

We show achievability by using a quantum version of the  binning technique.
Let $\varepsilon_i,\delta>0$ be arbitrarily small.
Define the average states,
\begin{align}
\sigma_{AB}^{(x)}&=
\sum_{y\in\mathcal{Y}} p_{Y|X}(y|x) 
\sigma_{AB}^{(x,y)}
\,,
\\
\sigma_{AC}^{(y)}&=
\sum_{x\in\mathcal{X}} p_{X|Y}(x|y) 
\sigma_{AC}^{(x,y)} \,,
\end{align}
and consider a spectral decomposition of the reduced states of Bob and Charlie,
\begin{align}
\sigma_{B}^{(x)}&=\sum_{z\in\mathcal{Z}}p_{Z|X}(z|x)\ketbra{\psi_{x,z}}\,,
\\
\sigma_{C}^{(y)}&=\sum_{w\in\mathcal{W}}p_{W|Y}(w|y)\ketbra{\phi_{y,w}}\,,
\end{align}
where $p_{Z|X}$ and $p_{W|Y}$ are conditional probability distributions, and 
$\{\ket{\psi_{x,z}}\}_{z}\ensuremath{,}\{\ket{\phi_{x,y}}\}_{w}$
are orthonormal 
bases for $\mathcal{H}_B$, $\mathcal{H}_C$, respectively, for every given $x\in\mathcal{X}$ and $y\in\mathcal{Y}$. We can also assume that the different bases are orthogonal to each other by requiring that Bob and Charlie encode on a different Hilbert space for every value of $(x,y)$.

We use the type class definitions and notations in 
\cite[Chap. 14]{wilde2017quantum}.
In particular,
$T_\delta^{X^n}$ denotes the $\delta$-typical set  
with respect to $p_X$, and 
 $T_\delta^{Z^n|x^n}$ is the conditional $\delta$-typical set with respect to $p_{XZ}$, given $x^n\in T_\delta^{X^n}$.

\paragraph*{Classical Codebook Generation}
For every sequence 
$z^{n}\in\mathcal{Z}^{n}$,   assign an index $m_1(z^{n})$,  uniformly at random from $[2^{nQ_1}]$. 
A bin $\mathfrak{B}_1(m_1)$ is defined as the 
 subset of sequences in $\mathcal{Z}^{n}$ that are assigned the same index $m_1$, 
 for $m_1\in[2^{nQ_1}]$. The codebook is revealed to all parties.

\paragraph*{Encoding}
\begin{enumerate}[A)]
\item
Alice prepares 
$\omega_{A\bar{B} \bar{C}}^{\otimes n}$ locally, where $\bar{B}^n \bar{C}^n$ are her ancillas,  without any correlation with $X^n$ and $Y^n$ (see Remark \ref{Remark:Broadcast_no_correlation}).
She applies the  encoding channel
$\mathcal{E}^{(1)}_{\bar{B}^n\to M_1}\otimes 
\mathcal{E}^{(2)}_{\bar{C}^n\to M_2}$,
\begin{align}
\mathcal{E}_{\bar{B}^{n}\to M_{1}}^{(1)}(\rho_{1})&=\sum_{x^{n}\in\mathcal{X}^{n}}
p_X^{\otimes n}(x^n)
\sum_{z^{n}\in\mathcal{Z}^{n}}\bra{\psi_{x^{n},z^{n}}}\rho_{1}\ket{\psi_{x^{n},z^{n}}}\ketbra{m_{1}(z^{n})}\,,
\\
\mathcal{E}_{\bar{C}^{n}\to M_{2}}^{(2)}(\rho_{2})&=\sum_{y^{n}\in\mathcal{Y}^{n}}p_Y^{\otimes n}(y^n)\sum_{w^{n}\in\mathcal{W}^{n}}\bra{\phi_{y^{n},w^{n}}}\rho_{2}\ket{\phi_{y^{n},w^{n}}}\ketbra{m_{2}(w^{n})}\,,
\end{align}
for $\rho_1\in\Delta(\mathcal{H}_B^{\otimes n})$, $\rho_2\in\Delta(\mathcal{H}_C^{\otimes n})$,
and transmits $M_1$ and $M_2$ to Bob and Charlie, respectively.

\item
First, Bob applies the following encoding channel,
\begin{align}
\mathcal{F}^{(x^n)}_{M_1\to B^n}(\rho_{M_1})=
\sum_{m_1=1}^{2^{nQ_1}} \bra{m_1} \rho_{M_1} \ket{m_1}
\left(\frac{1}{\abs{T_{\delta}^{Z^{n}|x^{n}}\cap
\mathfrak{B}_1(m_1)}}\sum_{z^n\in T_{\delta}^{Z^{n}|x^{n}}\cap
\mathfrak{B}_1(m_1)
}
\ketbra{\psi_{x^n,z^n}} \right)
\label{Equation:Broadcast_Bob_F_n}
\end{align}

\item
Charlie's decoder is defined in a similar manner.

\end{enumerate}

\paragraph*{Error analysis}
Due to the code construction, it suffices to consider the individual errors of Bob and Charlie,
\begin{align}
&\frac{1}{2}\left\Vert \omega_{XAB}^{\otimes n}-
\left(\mathcal{F}_{X^n M_{1}\to X^n B^{n}}\circ\mathcal{E}_{\bar{B}^{n}\to M_{1}}^{(1)}\right)(\omega_{X}^{\otimes n}\otimes\omega_{A\bar{B}}^{\otimes n})\right\Vert _{1}
\,,
\\
&\frac{1}{2}\left\Vert \omega_{YAC}^{\otimes n}-\left(\mathcal{D}_{Y^n M_{2}\to Y^n C^{n}}\circ\mathcal{E}_{\bar{C}^{n}\to M_{2}}^{(2)}\right)(\omega_{Y}^{\otimes n}\otimes\omega_{A\bar{C}}^{\otimes n})\right\Vert _{1} \,,
\end{align}
respectively, where we use the short notation 
$\mathcal{E}_{\bar{B}^{n}\to M_{1}}^{(1)}\equiv 
\mathrm{id}_{X^n A^n}\otimes \mathcal{E}_{\bar{B}^{n}\to M_{1}}^{(1)}$, and similarly for the other encoding maps.

We now focus on Bob's error. 
Consider a given codebook $\mathscr{C}_1=\{m_1(z^n)\}$.
 Alice encodes $M_1$ by
\begin{align}
\mathcal{E}_{\bar{B}^{n}\to M_{1}}^{(1)}(
\omega_{A B}^{\otimes n}
)
&=
\sum_{\tilde{x}^{n}\in \mathcal{X}^n}p_X^{\otimes n}(\tilde{x}^n)\sum_{z^n\in\mathcal{Z}^{n}}
\bra{\psi_{\tilde{x}^{n},z^{n}}}\omega_{A B}^{\otimes n}\ket{\psi_{\tilde{x}^{n},z^{n}}}\ketbra{m_{1}(z^{n})}
\,,
\end{align}
where we use the short notation
$\ket{\psi}_{x^n,z^n}\equiv \bigotimes_{i=1}^n \ket{\psi}_{x_i,z_i}$.
By the weak law of large numbers, this state is 
$\varepsilon_1$-close in trace distance to 
\begin{align}
\rho^{(1)}_{A^n M_1}
&=
\sum_{\tilde{x}^{n}\in  T_\delta^{X^n}}
p_X^{\otimes n}(\tilde{x}^n)\sum_{z^n\in T_{\delta}^{Z^{n}|\tilde{x}^{n}}}
\bra{\psi_{\tilde{x}^{n},z^{n}}}\sigma_{A^n \bar{B}^{n}}^{(\tilde{x}^{n})}\ket{\psi_{\tilde{x}^{n},z^{n}}}\ketbra{m_{1}(z^{n})}
\nonumber\\
&=
\sum_{{x}^{n}\in  T_\delta^{X^n}}
p_X^{\otimes n}(x^n) \rho^{(1|x^n)}_{A^n M_1}
\,,
\intertext{for sufficiently large $n$, where we have defined}
\rho^{(1|x^n)}_{A^n M_1}&=
\sum_{z^n\in T_{\delta}^{Z^{n}|x^{n}}}
\bra{\psi_{{x}^{n},z^{n}}}\sigma_{A^n \bar{B}^{n}}^{({x}^{n})}\ket{\psi_{{x}^{n},z^{n}}}\ketbra{m_{1}(z^{n})}
\,.
\end{align}

Let $x^{n}\in T_\delta^{X^n}$.
After Bob encodes $B^n$, we have
\begin{align}
\mathcal{F}_{M_{1}\to B^{n}}^{(x^{n})}\left(\rho^{(1|x^n)}_{A^n M_1}\right)
=\sum_{z^n\in T_{\delta}^{Z^{n}|x^{n}}}
\bra{\psi_{x^{n},z^{n}}}\sigma_{A^n \bar{B}^{n}}^{(x^{n})}\ket{\psi_{{x}^{n},z^{n}}}
\mathcal{F}_{M_{1}\to B^{n}}^{(x^{n})}(
\ketbra{m_{1}(z^{n})}
)
\,.
\label{Equation:Encoding_then_decoding_n}
\end{align}
By the definition of Bob's encoding channel, $\mathcal{F}_{M_{1}\to B^{n}}^{(x^n)}$, in \eqref{Equation:Broadcast_Bob_F_n}, 
\begin{align}
\mathcal{F}_{M_{1}\to B^{n}}^{(x^{n})}\left(\ketbra{m_{1}(z^{n})}\right)
=\frac{1}{\abs{T_{\delta}^{Z^{n}|x^{n}}\cap\mathfrak{B}_{1}(m_{1}(z^{n}))}}\sum_{\tilde{z}^{n}\in T_{\delta}^{Z^{n}|x^{n}}\cap\mathfrak{B}_{1}(m_{1}(z^{n}))}\ketbra{\psi_{x^{n},\tilde{z}}}
\,.
\label{Equation:Error_F}
\end{align}
Substituting in 
\eqref{Equation:Encoding_then_decoding_n} yields
\begin{align}
\mathcal{F}_{M_{1}\to B^{n}}^{(x^{n})}\left(\rho^{(1|x^n)}_{A^n M_1}\right)
= %
&\sum_{z^{n}\in T_{\delta}^{Z^{n}|x^n}}
\bra{\psi_{x^{n},z^{n}}}\sigma_{A^n \bar{B}^{n}}^{(x^{n})}\ket{\psi_{{x}^{n},z^{n}}}
\nonumber
\\
&\otimes \left[\frac{1}{\abs{T_{\delta}^{Z^{n}|x^{n}}\cap\mathfrak{B}_{1}(m_{1}(z^{n}))}}\sum_{\tilde{z}^{n}\in T_{\delta}^{Z^{n}|x^{n}}\cap\mathfrak{B}_{1}(m_{1}(z^{n}))}\ketbra{\psi_{x^{n},\tilde{z}^{n}}}\right]
\,.
\label{Equation:Broadcast_final_Bob}
\end{align}

Based on the classical result \cite[Chapter 10.3]{ElGamalKim:11b}, the random codebook $\mathscr{C}_1$ satisfies that %
\begin{align}
\Pr_{\mathscr{C}_1}\left(\exists \tilde{z}^n\in T_{\delta}^{Z^{n}|x^{n}}\cap \mathfrak{B}_{1}(m_{1}(z^{n})) :
\tilde{z}^n\neq z^n \right)\leq \varepsilon_2
\end{align}
 given $z^{n}\in T_{\delta}^{Z^{n}|x^n}$, for sufficiently large $n$, provided that the codebook size is at least $2^{n(H(Z|X)+\varepsilon_3)}$, where $H(Z|X)$ denotes the classical conditional entropy. As  $\abs{\mathscr{C}_1}=2^{nQ_1}$, this holds if
\begin{align}
Q_1&> H(Z|X)+\varepsilon_3
\nonumber\\
&= H(B|X)_\omega+\varepsilon_3 \,.
\end{align}
Observe that if the summation set in \eqref{Equation:Broadcast_final_Bob}, $T_{\delta}^{Z^{n}|x^{n}}\cap\mathfrak{B}_{1}(m_{1}(z^{n}))$,  consists of the sequence $z^n$ alone, then 
the overall state in \eqref{Equation:Broadcast_final_Bob} is identical to the post-measurement state after a typical subspace measurement on $B^n$, with respect to the conditional $\delta$-typical set
$T_\delta^{Z^n|x^n}$. Based on the gentle measurement lemma \cite{796385}, this state is $\varepsilon_4$-close to $\sigma_{AB}^{(x^n)}$, for sufficiently large $n$.

Therefore, by the triangle inequality and total expectation formula,
\begin{align}
&\left\| \omega_{XAB}^{\otimes n} - \mathbb{E}_{\mathscr{C}_{1}}\left( \mathcal{F}_{X^n M_{1}\to X^n B^{n}} \circ \mathcal{E}_{\bar{B}^{n}\to M_{1}}^{(1)} \right) \left(\omega_X^{\otimes n}
\otimes \omega_{A\bar{B}}^{\otimes n} \right) \right\|_{1}
\nonumber\\
&\leq
\sum_{x^n\in\mathcal{X}^n}p_X^{\otimes n}(x^n) \cdot \mathbb{E}_{\mathscr{C}_{1}} \left\| 
\sigma_{A^n B^n}^{(x^n)}
- \left( \mathcal{F}_{M_{1}\to B^{n}}^{(x^{n})} \circ \mathcal{E}_{\bar{B}^{n}\to M_{1}}^{(1)} \right) \left( \sigma_{A^n B^n}^{(x^n)} \right) \right\|_{1}
\nonumber 
\\
&\leq \varepsilon_1+\varepsilon_2+\varepsilon_4 \,.
\end{align}
By symmetry, Charlie's error tends to zero as well, provided that
$Q_2\geq H(C|Y)_\omega+\varepsilon_5$.
Since the total error vanish, when averaged over the class of binning codebooks, it follows that there exists a deterministic codebook with the same property.
The achievability proof follows by taking
$n\to\infty$ and then $\varepsilon_j,\delta\to 0$. 

The converse proof follows the lines of \cite{Yard_Devetak_2009}, and it is thus omitted. This completes the proof of Theorem~\ref{Theorem_Broadcast}
for the broadcast network.

\section{Multiple-Access  Analysis}
\label{Multiple_access_network_analysis}
 We prove the rate characterization in Theorem~\ref{MAC_theorem}. Consider the multiple-access network in Figure~\ref{Multiple_access_network}.
As explained in Remark~\ref{MAC_product_state}, coordination in the multiple-access network is only possible if
there exists an isometry $V:\mathcal{H}_C\to \mathcal{H}_{C_1}\otimes \mathcal{H}_{C_2}$ such that
\begin{align}
(\identity\otimes V)\ket{\omega_{A B C }}=
\ket{\phi_{AC_1}}\otimes \ket{\chi_{BC_2}}
\label{Equation:MAC_Isometry_Converse}
\end{align}
where $\ket{\phi_{AC_1}}$ and $\ket{\chi_{BC_2}}$ are purifications of $\omega_A$ and $\omega_B$, respectively.
For this reason, Theorem~\ref{MAC_theorem} assumes that this property holds. Furthermore, since $\ket{\phi_{AC_1}}$ and $\ket{\chi_{BC_2}}$  purify $\omega_A$ and $\omega_B$, respectively, we have $H(C_1)_\phi=H(A)_\phi=H(A)_\omega$ and
$H(C_2)_\chi=H(B)_\chi=H(B)_\omega$. Thus, it suffices to show that $(Q_1,Q_2)$ is achievable if and only if
\begin{align}
Q_1&\geq H(C_1)_\phi \,,
\\
Q_2&\geq H(C_2)_\chi \,.
\end{align}

The achievability proof  follows from the Schumacher compression protocol \cite{schumacher1995quantum} \cite[chap. 18]{wilde2017quantum} in a straightforward manner.
Alice and Bob prepare $\phi_{AC_1}^{\otimes n}$ and $\chi_{BC_2}^{\otimes n}$, respectively.
Then, they send $C_1^n$ and $C_2^n$ using the Schumacher compression protocol, and finally, Charlie 
applies the isometry $(V^{\dagger})^{\otimes n}$ in order to simulate $\omega_{ABC}^{\otimes n}$ (see \eqref{Equation:MAC_Isometry_Converse}). The details are omitted.

It remains to show the converse part. 
Recall that in the multiple-access network, Alice and Bob each applies their respective encoding map, $\mathcal{E}_{A^{n}\to A^{n}M_{1}}$ and $\mathcal{F}_{B^{n}\to B^{n}M_{2}}$, and send the quantum descriptions $M_1$ and $M_2$. Then,  Charlie encodes by $\mathcal{D}_{M_{1}M_{2}\to C^{n}}$. 
The protocol can be described through the following relations:
\begin{align}
\rho_{A^{n}M_{1}}^{(1)}&=\mathcal{E}_{A^{n}\to A^{n}M_{1}}(\omega_{A}^{\otimes n})\,,\; %
\rho_{B^{n}M_{2}}^{(2)}=\mathcal{F}_{B^{n}\to B^{n}M_{2}}(\omega_{B}^{\otimes n})\,,
\label{MAC_converse_rho_2} 
\\
\widehat{\rho}_{A^{n}B^{n}C^{n} }
&=\left(\mathrm{id}_{A^{n}B^{n}}\otimes\mathcal{D}_{M_{1}M_{2}\to C^{n}}\right)(\rho_{A^{n}M_{1}}^{(1)}\otimes\rho^{(2)}_{B^{n}M_{2}}) \,.
\label{MAC_converse_rho_hat} 
\end{align} 
Let $\left(Q_{1},Q_{2}\right)$ be an achievable rate pair for coordination in the multiple-access network in Figure~\ref{Multiple_access_network}. Then,  there exists a sequence of $\ensuremath{(2^{nQ_{1}},2^{nQ_{2}},n)}$ coordination codes such that
\begin{align} 
\left\Vert \widehat{\rho}_{A^{n}B^{n}C^{n}}-\omega_{ABC}^{\otimes n}\right\Vert _{1}\leq\varepsilon_n
\label{MAC_converse_error} 
\end{align}
tends to zero as $n\to \infty$.
Applying the isometry $V^{\otimes n}$ yields 
\begin{align} 
\left\Vert \widehat{\sigma}_{A^{n}B^{n}C_1^{n} C_2^n}-\phi_{AC_1}^{\otimes n}\otimes \chi_{BC_2}^{\otimes n}\right\Vert _{1}\leq\varepsilon_n \,,\label{Equation:MAC_converse_error} 
\end{align}
by \eqref{Equation:MAC_Isometry_Converse},
where 
\begin{align}
\widehat{\sigma}_{A^{n}B^{n}C_1^{n} C_2^n}
=
(
\identity_{AB}\otimes V)^{\otimes n}
\widehat{\rho}_{A^{n}B^{n}C^{n}} 
(\identity_{A B }\otimes V^{\dagger})^{\otimes n}
\,.
\end{align}
It thus follows that
\begin{align} 
\left\Vert \widehat{\sigma}_{A^{n}C_1^{n} }-\phi_{AC_1}^{\otimes n}\right\Vert _{1}\leq\varepsilon_n 
\intertext{and}
\left\Vert \widehat{\sigma}_{B^{n}C_2^{n} }-\chi_{BC_2}^{\otimes n}\right\Vert _{1}\leq\varepsilon_n \,.
\label{Equation:Multiple_access_converse_error_3} 
\end{align}

Now, Alice's communication rate is bounded by
\begin{align}
2nQ_{1}	
&\overset{(a)}{\geq} I(M_{1};A^{n}|M_2)_{\rho^{(1)}\otimes \rho^{(2)}}
\nonumber\\
&\overset{(b)}{=} I(M_{1}M_{2};A^{n})_{\rho^{(1)}\otimes \rho^{(2)}}
\nonumber\\
&\overset{(c)}{\geq} 
I(C^n;A^{n})_{\widehat{\rho}}
\nonumber\\
&\overset{(d)}{=} 
I(C_1^n C_2^n;A^{n})_{\widehat{\sigma}}
\nonumber\\
&\overset{(e)}{\geq} 
I(C_1^n C_2^n;A^{n})_{\omega}-n\alpha_n
\nonumber \\
&\overset{(f)}{=} 
2n H(C_1)_{\phi}-n\alpha_n
\nonumber \\
&\overset{(g)}{=} 
2n H(A)_{\omega}-n\alpha_n\,,
\end{align}
where 
$(a)$
holds because $M_1$ is of dimension
$2^{nQ_1}$, 
$(b)$ since $I(M_{2};{A}^n)_{\rho^{(1)}\otimes \rho^{(2)}}=0$,
$(c)$ follows from the data processing inequality,
$(d)$ holds since the von Neumann entropy is isometrically invariant, 
$(e)$  by the AFW inequality \cite{AFW_Winter_2016}
, $(f)$ since the mutual information is calculated with respect to the  product state $\ket{\phi_{AC_1}}^{\otimes n}\otimes \ket{\chi_{BC_2}}^{\otimes n}$, and $(g)$ holds since $\ket{\phi_{AC_1}}$ is a purification of $\omega_A$.
The bound on Bob's communication rate follows by symmetry.
This completes the proof of Theorem~\ref{MAC_theorem}.
\qed

\section{Summary and discussion}
\label{Section:SummarynDiscussion}
 \subsection{Summary}
We study coordination in 
three network models with quantum links: 1) a cascade network simulating quantum states with limited communication and entanglement assistance, 2) a quantum linked broadcast network simulating a joint quantum state  with classical side information, and 3) a multiple-access network, generating  entanglement between each  sender and the receiver.  
We observe that the network topology  dictates the type of states that can be simulated. 

The results are relevant for various applications, where the network nodes could represent 
computers performing a joint computation task \cite{6G_quantum_2022,pereg2022classical}, 
or players in a nonlocal game \cite{Seshadri:23p,pereg2023multiple} as we illustrated in the  broadcast network with quantum links, in which we establish the optimal rates required to achieve a certain quantum correlation to win a game at a desired probability.

\subsection{General Coordination Problem}
The coordination  problem in its general form can be presented as follows. Given a network consisting of $N$ nodes, the nodes need to cooperate using limited  resources to asymptotically achieve a joint state that is arbitrarily close in trace distance to a desired joint state $\omega_{A_1^n \dots A_N^n}$. 
The objective of the coordination task is to find the optimal resources needed for simulating the correlation manifested in the desired joint state. We represent each user in the coordination network by a node, each node is assigned an index $j\in[N]$ and has access to a quantum system $A_j^n$. 
The users may communicate using classical or quantum communication links. In this paper we considered one-way links, in general they can be two-way links. We denote the classical message transmitted from Node~$j$ to Node~$k$ by $m_{j,k}$ with a rate $R_{j,k}$, similarly, a quantum message (description) is denoted by $M_{j,k}$, transmitted at a rate $Q_{j,k}$. In addition to the limited communication links, the users have access to additional resources such as common randomness (CR) at a rate $R_{0}$, entanglement assistance at a rate $E_{j,k}$, and side information for Node~{j} which can be either classical or quantum. 
Each user performs an encoding operation $\mathcal{E}_j$ on his system $A_j$, where he applies the encoding operation on the resources available in his possession. Receiving side information, incoming classical messages $m_{l,j}$ and quantum descriptions $M_{l,j}$, using his share of CR and entanglement assistance, he prepares the state of his system $\rho_{A_j}$, in addition to the  outgoing message $m_{j,k}$, and the joint state of the quantum descriptions $M_{j,k}$. After $n$ time steps, the network ends up in a joint state $\widehat{\rho}_{R^n A_1^n \dots A_N^n}$, where $R^n$ is a reference system. The users would like this state to be close to 
$\omega_{R A_1 \dots A_N}^{\otimes n}$.
Below, we provide generic definitions for the coordination problem.

\begin{definition}
    A rate tuple $\left( 
    R_{0}
    ,\left\{R_{j,k}\right\}
    ,\left\{E_{j,k}\right\}
    ,\left\{Q_{j,k}\right\}
    \right)_{(j,k)\in[N]}$
    is achievable for the simulation of 
$\omega_{A_1 \dots A_N}$ in the  network if for every $\varepsilon,\delta>0$ and a sufficiently large $n$,
there exists a  
$\left( 
    2^{nR_{0}}
    ,\left\{2^{nR_{j,k}}\right\}
    ,\left\{2^{nE_{j,k}}\right\}
    ,\left\{2^{nQ_{j,k}}\right\}
    ,n\right)_{j,k\in[N]}$
code that achieves
\begin{align}
\norm{\widehat{\rho}_{R^n A_1^n \dots A_N^n}-\omega_{R A_1 \dots A_N}^{\otimes n}}_1\leq \varepsilon \,.
\end{align}
\end{definition}

\begin{definition}
The coordination
capacity region  with respect to a desired state $\omega_{A_1 \dots A_N}^{\otimes n}$, is the closure of the set of all achievable rate tuples. 
\end{definition}
This work can be viewed as a step forward in understanding coordination in a general network that may comprise either classical or quantum resources. 
While our results cover fundamental building blocks for quantum network coordination, we do not claim to have solved all network settings. 
The no-communication network was independently considered by George et al.  \cite{%
george2023one} (see also \cite{george2023one_arXiv,george2024reexamination}) in both the one-shot and asymptotic settings.
Another interesting direction for future work is the  characterization of network coordination  in the one-shot regime for 
the two-node, broadcast, multiple-access and cascade networks.

\subsection{Correlation resources}
We  study the effect of pre-shared correlation on network coordination, focusing on 
pre-shared entanglement with quantum links. 
George and Cheng \cite{george2024coherent} have recently considered a broadcast setting of state splitting, including classical links and entanglement assistance (see also \cite{cheng2023quantum}).
It would be interesting to further study coordination networks with classical links and pre-shared  entanglement at a limited rate. 
The general task of quantum network coordination can be viewed as a generalization of   channel/source simulation \cite{
channelsimulation2024,
berta2013entanglement,
wilde2018entanglement,
cheng2023quantum,george2023one,allah2024quantum,chitambar2022communication},
resolvability and soft covering \cite{wyner1975common,winter2001compression,hayashi2016quantum,Bloch_resolvability_2019},
state merging \cite{bjelakovic2013universal,horodecki2007quantum},
state redistribution \cite{Yard_Devetak_2009,
berta2016smooth
},
entanglement dilution \cite{hayden2003communication,harrow2004tight,kumagai2013entanglement,
Salek_2022_Winter},
randomness extraction \cite{6670761,
tahmasbi2020steganography},
source coding \cite{goldfeld2014ahlswede,kramer2001quantum,Compressing_mixed_state_sources_2002, 
Rate_limited_source_coding_2023}, 
and many others (see Table~\ref{Table: Related problems}).

%


\subsection{Quantum repeaters}
\label{Subsection: Quantum repeaters} 
\begin{figure}[t]
\center
\includegraphics[scale=0.75,trim={5.3cm 0 5.5cm 0}]
{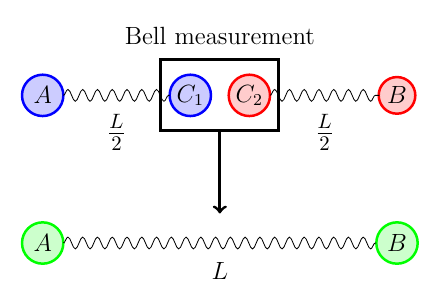} 
\caption{A quantum repeater link \cite{van2020extending}.
Initially, systems $C_1$ and $C_2$ belong to Charlie. $C_1$ is maximally entangled with Alice's system $A$ (the blue pair), while $C_2$ is maximally entangled with Bob's system $B$ (the red pair). Both pairs of entangled systems are at $\frac{L}{2}$ distance apart. When Charlie performs a local Bell measurement on his systems, causing entanglement swapping. Such that at the end of the protocol, systems $A$ and $B$ are maximally entangled (the green pair), which are at a distance $L$ from each other, creating a longer distance entanglement which can be exploited further .}
\label{Figure: Repeaters}
\end{figure}
Quantum communication  relies on
the transmission of quantum signals over long distances \cite{munro2015inside}. Unfortunately, long-range communication is limited by attenuation and loss
\cite{pereg2023communication}. 
%
While classical systems can overcome such losses with straightforward signal amplification, this solution is not viable for quantum signals due to the no-cloning theorem, which prohibits the 
replication of quantum states.
Therefore, alternative methods are required to address these limitations. 

Quantum repeaters offer a promising solution by enabling the generation of entanglement between distant network nodes. 
The process begins by  
creating maximally entangled pairs between neighboring nodes, 
and then extending this entanglement to 
a longer range \cite{briegel1998quantum}. 
This long-range entanglement facilitates quantum teleportation, which enables the transmission of quantum information between the sender and receiver.
 Quantum repeaters are expected to play a central role in the future quantum Internet
\cite{goodenough2021optimizing}, and recent advancements in their implementation have been explored across various experimental platforms
\cite{wilksen2024gate,neuwirth2021quantum,
bergerhoff2024quantum,dhara2022multiplexed,
mizuochi2024quantum,azuma2023quantum,rodgers2021materials}.

In the simplest description of a quantum repeater, the process begins with using quantum communication and entanglement distillation to
prepare two pairs of qubits at maximally entangled states, namely, $\ket{\Phi_{AC_1}}$ between the sender and the repeater, and $\ket{\Phi_{BC_2}}$ between the repeater
and the receiver. At the next stage, the repeater performs a Bell measurement on $C_1$ and $C_2$, thus swapping the entanglement such that $A$ and $B$
are now entangled at a distance twice that of the initial entangled pairs.
See Figure~\ref{Figure: Repeaters}.
Different information-theoretic models for quantum repeaters can be found in \cite{pereg2024quantum,PeregDeppeBoche:21p2,gyongyosi2012private,shi2012lower,gyongyosi2014reliable,pirandola2016capacities,ghalaii2020capacity}. 

Coordination is highly relevant for quantum repeaters. 
For instance,
consider the simulation of a product of two maximally entangled pairs of qubits, $\ket{\omega_{ABC_1 C_2}}=\ket{\Phi_{AC_1}}\otimes \ket{\Phi_{BC_2}}$ over the multiple-access network in Figure~\ref{Multiple_access_network}. The coordination capacity region for this setup is $\mathcal{Q}_{\text{MAC}}(\omega)=\{(Q_1,Q_2):Q_i\geq 1\}$, which reflects the requirement of sending a single qubit from the repeater (Charlie), to each user (Alice and Bob), to prepare the two entangled pairs. See Remark~\ref{Remark: repeaters}. 
As described above, a quantum repeater can transform this correlation into maximal entanglement between $A$ and $B$.
More generally, a quantum repeater can be adapted to simulate a wide range of quantum correlations, with the required communication rates being determined by the results of Theorem~\ref{MAC_theorem}.
Thus, the coordination features of a multiple-access network are not only relevant but also beneficial in the context of quantum repeater networks.
The findings from Theorem~\ref{MAC_theorem} highlight the optimal communication rates necessary for generating entangled states beyond just maximally entangled pairs, broadening the potential applications of quantum repeaters.



%

\section*{Acknowledgements}
The authors would like to thank Ian George (NUS), 
Eric Chitambar (UI 
Urbana-Champaign), and
Marius Junge (UI 
Urbana-Champaign)
for useful discussions during the conference BIID 2024, 
supported by NSF Grant n. 2409823.

HN and UP are
grateful to the 
Helen Diller Quantum Center
at
the Technion
for supporting this research and acknowledge support
from
the  Israel Science Foundation (ISF), Grants 939/23 and 2691/23, German-Israeli Project Cooperation (DIP), Grant
2032991, and  Nevet Program of the Helen Diller Quantum Center, Grant 	2033613.
UP was also supported by the Junior Faculty Program for Quantum Science and Technology through
of the Planning and Budgeting Committee of the Council for Higher Education of Israel (VATAT),
Grant 86636903, and the Chaya Career Advancement Chair, Grant 8776026.

\bibliography{references_2}
\end{document}